\documentstyle[12pt,aps,axodraw,epsfig]{revtex}
\begin{document}

\def\up{\partial^{\mu}}
\def\down{\partial_{\mu}}
\def\ZEROVEC{\underline{0}}
\def\ds{\rlap/\partial}
\def\epsilons{\rlap/\epsilon}
\def\ks{\rlap/k}
\def\ps{\rlap/p}
\def\qs{\rlap/q}
\def\O{{\cal O}}
\def\gapprox{\stackrel{>}{_\sim}}
\def\lapprox{\stackrel{<}{_\sim}}

\title{
\begin{flushright}
\vspace{-1cm}
{\normalsize MC/TH 97/06}
\vspace{1cm}
\end{flushright}
Meson properties in an extended nonlocal NJL model
\vskip 10pt}
\author{Robert S. Plant and Michael C. Birse}
\address{Theoretical Physics Group, Department of Physics and Astronomy,\\
University of Manchester, Manchester, M13 9PL, U.K.\\}
\maketitle
\vskip 20pt

\setlength{\unitlength}{1pt}

\begin{abstract}
We consider a nonlocal version of the NJL model, based on a separable
quark-quark interaction. The interaction is extended to include terms that
bind vector and axial-vector mesons. The nonlocality means that no further
regulator is required. Moreover the model is able to confine the quarks by
generating a quark propagator without poles at real energies. Working in the
ladder approximation, we calculate amplitudes in Euclidean space and discuss
features of their continuation to Minkowski energies. Conserved currents are
constructed and we demonstrate their consistency with various Ward identities.
Various meson masses are calculated, along with their strong and
electromagnetic decay amplitudes. We also calculate the electromagnetic form 
factor of the pion, as well as form factors associated with the processes
$\gamma\gamma^*\rightarrow\pi^0$ and $\omega\rightarrow\pi^0\gamma^*$. The
results are found to lead to a satisfactory phenomenology and lend some
dynamical support to the idea of vector-meson dominance.
\end{abstract}
\vskip 20pt

\section{Introduction}
The Nambu--Jona-Lasinio (NJL) model\cite{NJL,NJLrevs,ENJLrev} has long been
used as a starting point for the description of light mesonic states as
fermion-antifermion composites. It shares several features with QCD, notably a
dynamically-broken chiral symmetry, with the pions as approximate Goldstone
bosons. The model is based on fermionic fields interacting through a local,
chirally-invariant, four-point vertex. The local nature of the interaction
produces a great simplification of the Schwinger--Dyson and Bethe--Salpeter
equations. The main drawbacks of the model are, however, direct consequences
of this locality. Specifically, they are that the loop integrals diverge (and
so must somehow be regulated) and that the model is non-confining.

Although the NJL model does contain regularization-independent
information\cite{Bij94a,Bij94b} and the results with various regularization
schemes have been found to be qualitatively similar\cite{MRG90}, the choice of
any particular scheme lacks a sound physical motivation. A feature of
many regularization schemes is that as well as the form of the cut-off, a
particular momentum routing must be specified for loop diagrams with
two-or-more quark lines\cite{WG}. In practice a symmetric routing is often
implicitly taken in order to maintain Ward identities. The regularization
scheme must be specified yet further if one wishes to calculate beyond leading
order in the $1/N_c$ expansion, a new cut-off being required for meson
loops\cite{NBCRG,BP97}.

Another unsatisfactory aspect is that if low-energy theorems for anomalous
processes (for example $\pi^0 \to \gamma\gamma$) are to hold then the
anomalous diagrams must either be left unregulated\cite{SDAR} or else
additional terms must be added to the Lagrangian in order to respect the
anomalous Ward identities\cite{Bij94b}. Related problems occur in the
presence of vector interactions\cite{SDAR,ARWZ,AS93,AR96} if one attempts to
apply the regularization prescription to both the anomalous and non-anomalous
sectors.

Many attempts have been made to generalize the original NJL
model\cite{beyondNJL} to remove such unwanted features whilst retaining its 
successful phenomenological aspects\cite{NJLrevs,ENJLrev}. One promising 
approach, which provides some motivation for the model studied here, is 
suggested by instanton-liquid studies\cite{DP}. In that picture, the instantons
induce an effective four-quark vertex, which is nonlocal but of a separable 
form. The separable nature of the interaction retains as far as possible the 
simplifying features of a local model, with the nonlocality providing a cut-off
on all loop integrals. A similar class of model has a separable dependence on
the relative momentum of the $\overline{q}q$ pair and has been studied in
Refs.\cite{IBG,SBK}. Other models with simple interactions have been suggested
based on various types of gluonic field configurations within the QCD vacuum.
For example, Efimov and coworkers\cite{Efim} start with a constant (anti-)
self-dual background field in Euclidean space and base their four-quark
vertex on one-gluon exchange within such a background. Yet another recent
model\cite{LR96} used a four-quark vertex mediated by a random colour-matrix,
as an attempt to simulate a strongly-fluctuating background gluon field
(see also\cite{Rei91}). We should mention that there are also studies of QCD
Schwinger--Dyson equations based on one-gluon exchange forces between the
quarks, often using effective gluon propagators\cite{GCM,FR96} (also
see\cite{SDErev} and references therein).

Here we develop further the model proposed in Ref.~\cite{BB95}. This is based
on a nonlocal, separable, four-quark vertex. It is therefore similar to the
instanton-liquid model of Ref.~\cite{DP}, the differences being that more
general choices of the form factor and possible couplings are considered. The
particular choice of form factor used in this model can lead to quark
confinement, in the sense of a quark propagator without poles at real energies.
It also ensures convergence of the quark-loop integrals, unlike that chosen in
the separable model of Ref.~\cite{BK92}. Only the pions and their scalar 
partner were studied in Ref.~\cite{BB95}. In the spirit of the extended NJL
model~\cite{ENJLrev,Bij94a,Bij94b,SDAR,ENJL,ENJLboson,Pr94,KLVW,BJM88}, we 
incorporate other mesonic degrees of freedom, such as the vector mesons. 
Including these particles enables us to probe the role of the confinement 
mechanism, since they have masses of around twice a typical constituent-quark 
mass. In a model without confinement, and with an otherwise reasonable
constituent-quark mass of $\sim 300$ MeV, the $\rho$ meson would lie above the
$\overline{q}q$ threshold and could decay into free $\overline{q}q$ states.

As well as the meson spectrum, we calculate hadronic decays of these particles
and some electromagnetic processes, including $\rho\to e^+e^-$ and the pion
form factor. With regard to the electromagnetic processes, we also examine
whether there is any support for the concept of vector-meson dominance within
a model of this type.

In our present calculations, we work at leading-order in the $1/N_c$ expansion.
Our results are thus comparable to those of most NJL-model and Schwinger-Dyson
approaches. At next-to-leading order, one has to include Fock terms arising
from the four-point interaction and meson-loop contributions. Calculations at
this order in the NJL model\cite{NBCRG,BP97} indicate that such corrections
can be significant, especially in the scalar-isoscalar channel. This may 
explain some of the failures of the model to give a good description of
amplitudes involving this channel.

This paper is organized as follows. In Sec.\ \ref{s2} we define the extended
model and give the forms of the corresponding vector and axial currents. In 
Sec.\ \ref{s3} we present the forms of the quark and meson propagators and
describe the means of coupling to external currents. The determination of the 
model parameters is described in Sec.\ \ref{s4}, and results for hadronic and 
electromagnetic decays of mesons are presented in Secs.\ \ref{s5} and \ref{s6} 
respectively. 

\section{The model}\label{s2}
Formally at least, one can imagine integrating out gluonic degrees of freedom
to leave an effective action for QCD expressed in terms of quark fields only. 
As in the usual NJL model, we keep only two-body forces between the quarks,
described by four-quark interaction vertices. Indeed, at leading order in
$1/N_c$, all six-quark and higher interactions could be absorbed into
effective couplings for the four-quark terms, by replacing extra
$\overline{\psi}\Gamma \psi$ factors with their vacuum expectation values.
This is just the procedure followed in the three-flavour extended NJL
model\cite{KLVW,BJM88} with a six-quark, U(1)$_A$-breaking 't Hooft
determinant\cite{tH76}. In the present work, however, we specialize to
two-flavours with isospin symmetry. The action may be written as
\begin{equation}
S= \int\! d^4\!x\, \overline{\psi}(x) (i \ds - m_c) \psi (x) 
+ \sum_i\int \prod_n d^4\! x_n\, H_{i} (x_1,x_2,x_3,x_4)\, \overline{\psi}(x_1)
\Gamma^\alpha_i \psi (x_3) \overline{\psi}(x_2) \Gamma_{i\alpha} \psi (x_4) . 
\label{4q action}
\end{equation}

Imposing SU(2)$_R\times$SU(2)$_L\times$U(1)$_V$ symmetry requires that certain
of the possible Dirac and isospin structures appear in the combinations:
\begin{eqnarray}
H_1(1 \otimes 1 + i\gamma_5 \tau^a \otimes i\gamma_5 \tau^a), &&\qquad 
H_2(\gamma^\mu \tau^a \otimes \gamma_\mu \tau^a + \gamma^\mu \gamma_5 \tau^a 
\otimes \gamma_\mu \gamma_5 \tau^a), \nonumber\\
H_5(\tau^a \otimes \tau^a + i\gamma_5 \otimes i\gamma_5), &&\qquad 
H_6(\sigma_{\mu\nu} \otimes \sigma^{\mu\nu} - \sigma_{\mu\nu} \tau^a \otimes 
\sigma^{\mu\nu} \tau^a), \label{chiral relations}  
\end{eqnarray}
whilst the strengths of the following interactions are unconstrained by
symmetry considerations:
\begin{equation}
H_3(\gamma^\mu \otimes \gamma_\mu),\qquad H_4(\gamma^\mu \gamma_5 \otimes 
\gamma_\mu \gamma_5). \label{indep interact}
\end{equation}

A variety of models of the above type can be found in the literature, differing
in the ansatz chosen for $\{ H_i (x_1,x_2,x_3,x_4) \}$. The original NJL model,
for instance, has $H_1 \sim \int d^4 x \prod_n \delta (x - x_n)$  and a 
constant coupling strength, whereas one-gluon exchange models take $H_i \sim 
\delta (x_1 - x_3) \delta (x_2 - x_4) D(x_1 - x_2)$. Our own approach is 
motivated in part by the instanton-liquid model\cite{DP}. Within the zero-mode 
approximation to that picture, there is a 2$N_f$-quark interaction, which is of
separable form,
\begin{equation}
H_i (p_1 ,p_2 ,p_3 ,p_4 )= {\textstyle\frac{1}{2}} (2\pi)^4 G_i f(p_1) f(p_2) 
f(p_3) f(p_4) \delta (p_1 + p_2 - p_3 - p_4) . \label{separable}
\end{equation}
In the model of Ref.~\cite{DP} the function $f(p)$ is of a particular form and
for two flavours of quarks the relation $G_1=-G_5$ follows from the form of
the 't Hooft determinant\cite{tH76}. An interaction of tensor character is also
present but is $1/N_c$ suppressed.

The model studied here is similar to that of Dyakonov and Petrov\cite{DP}, in
that it is based on an interaction with the separable form (\ref{separable}).
However, we shall adopt a more phenomenological attitude towards the
form-factor $f(p)$ and the allowed couplings. Only interactions in the
colour-singlet channels are considered in the present study. The $G_1$ coupling 
(in the ladder approximation) produces the pions and their isoscalar scalar
partner, $\sigma$. The couplings in the spin-1 channels $G_2$, $G_3$ and $G_4$
are responsible for the $\rho$, $a_1$, $\omega$ and $f_1$ mesons. The $G_5$
coupling allows the model to describe an isovector scalar and an isoscalar
pseudoscalar meson. The lowest-lying meson with quantum numbers corresponding
to the former is $a_0(980)$, whilst the latter is a non-strange state with the
quantum numbers of the $\eta$ and $\eta^\prime$, which we shall refer to as
$\eta^\star$.

We shall not consider the possible tensor interactions, described by the
coupling $G_6$. As discussed in Ref.~\cite{KLVW}, these can contribute in the
vector channels, where they give rise to anomalous magnetic-moment couplings of
the vector mesons to constituent quarks. In the absence of any strong 
phenomenological need for such couplings we choose not to include them.

For the sake of simplicity, all of the possible independent interactions
(\ref{chiral relations},\ref{indep interact}) are assumed to contain the same
form factor and to differ only in the constant coupling strengths, $G_i$. As
in Ref.\cite{BB95}, we take the form factor to be Gaussian in Euclidean space:
\begin{equation}
f(p_E)=\exp(-p_E^2 / \Lambda^2 ).
\end{equation}
This choice was shown to be able to give quark confinement. We have in fact
examined the possibility of taking a different $\Lambda$ for each of the
independent couplings. Since this does not lead to any very significant
effects, the results are not presented here. This is because the main features
are dominated by the form of the quark self-energy which, in the ladder
approximation, depends only on the $G_1$ interaction.

The usual, local expressions for the vector and axial currents do not satisfy
the correct continuity equations when one uses the equations of motion derived
from the action (\ref{4q action}--\ref{separable}). In order to obtain symmetry
currents with the same divergences as the corresponding local currents in QCD,
and hence to maintain the corresponding Ward identities, one has to introduce
additional nonlocal terms in the currents. A Noether-like method of
construction for these nonlocal pieces was developed in Ref.~\cite{BB95}. As
a part of that procedure the following identity was used,
\begin{equation}
\delta (x-x_1) - \delta (x-x_2) = \int_0^1 d\lambda \frac{dz^\mu}{d\lambda}
\down \delta (x-z) ,
\end{equation}
$z(\lambda)$ being some arbitrary path from $x_1$ to $x_2$. The divergence
requirement determines the longitudinal component of a current which is,
therefore, a path-independent object. In Ref.\cite{BB95} the choice of path was
irrelevant since the authors were interested only in the longitudinal component
of the axial current, in order to determine the pion decay constant. The 
transverse part of the current, however, is sensitive to the particular choice 
of $z(\lambda)$. Indeed, ambiguity in the transverse current is a feature of 
any method used to construct a (partially) conserved current corresponding to a
nonlocal action. When one wishes to consider electromagnetic processes, as in 
the present work, it is necessary to assume some form for the transverse 
current. This assumption forms an additional part of the specification of the 
model. In our present calculations, we adopt the straight line 
ansatz\cite{BB95},
\begin{equation}
z(\lambda)=(1-\lambda)x_1 + \lambda x_2 \label{linpath},
\end{equation}
since this choice respects both Lorentz and translational invariance. In fact 
a number of the electromagnetic observables calculated in Sec.\ref{s6} are 
dominated by the local piece of the vector current and so should not be 
very sensitive to the choice of path.

The nonlocal interactions in the model defined by (\ref{4q
action}--\ref{separable}) give rise to various nonlocal terms in the currents,
whose momentum-space forms are presented below. (Note that where momentum
derivatives with respect to $p_i \pm p_j$ occur, then the combination $p_i \mp
p_j$ is understood to be held fixed.) For the isoscalar vector current, the
nonlocal pieces all have the structure, 
\begin{eqnarray}
J^\mu_{(I)}&=& \frac{1}{(2\pi)^{12}} \sum_i G_i \int \prod_n d^4 
p_n ~\, \overline{\psi} (p_1) \Gamma_i^\alpha \psi (p_3) \overline{\psi} (p_2) 
\Omega_{i\alpha} \psi (p_4) ~\, \delta(p_1 + p_2 + q - p_3 - p_4)\nonumber\\
&&\qquad\qquad\times\int_0^1 d\lambda\, f(p_2)f(p_4) \frac{\partial}{\partial 
(p_1 + p_3)_\mu}f(p_1 + \lambda q) f(p_3 - q + \lambda q) , \label{I}
\end{eqnarray}
which we refer to as type I. The sum over $G_i (\Gamma_i^\alpha \otimes
\Omega_{i\alpha})$ in (\ref{I}) runs over the same combinations of couplings
and Dirac and isospin matrices as those found in the action, (\ref{chiral
relations},\ref{indep interact}).

The isovector vector current also has nonlocal contributions of this type-I
structure (\ref{I}). In this case the isospin and Dirac matrices appear in the
combinations:
\begin{eqnarray}
G_1 (\tau^a \otimes 1 + i\gamma_5\otimes i\gamma_5 \tau^a),&&\qquad 
G_2 (\gamma^\nu\otimes \gamma_\nu \tau^a + \gamma^\nu \gamma_5 \otimes 
\gamma_\nu \gamma_5 \tau^a),\nonumber\\
G_3 (\gamma^\nu \tau^a \otimes \gamma_\nu),&&\qquad G_4 (\gamma^\nu \gamma_5 
\tau^a \otimes \gamma_\nu \gamma_5), \qquad G_5 (1 \otimes \tau^a + i\gamma_5
\tau^a \otimes i\gamma_5) . \label{vec I}
\end{eqnarray}
Another type of nonlocal structure also arises in this current,
\begin{eqnarray}
J^\mu_{(II)}&=&\frac{i \epsilon^{abc}}{2(2\pi)^{12}} \sum_i G_i 
\int \prod_n d^4 p_n ~\, \overline{\psi} (p_1) \Gamma_i^\alpha \tau^b \psi 
(p_3) \overline{\psi} (p_2) \Omega_{i\alpha} \tau^c \psi (p_4) ~\, \delta(p_1 +
p_2 + q - p_3 - p_4) \nonumber\\
&&\qquad\qquad\times\int_0^1 d\lambda \left[ f(p_1)f(p_2) \frac{\partial}
{\partial (p_3 - p_4)_\mu}f(p_3 - q + \lambda q) f(p_4 - \lambda q) \right. 
\nonumber\\
&&\qquad\qquad\qquad\qquad\left. -f(p_3)f(p_4)\frac{\partial}{\partial 
(p_1 - p_2)_\mu} f(p_1 + q - \lambda q) f(p_2 + \lambda q) \right] . \label{II}
\end{eqnarray}
This type-II structure contributes in those interaction channels corresponding
to isovector states. The Dirac matrices appear in the combinations:
\begin{equation}
G_1 (i\gamma_5 \otimes i\gamma_5), \qquad G_2 (\gamma^\nu \otimes \gamma_\nu
+ \gamma^\nu \gamma_5 \otimes \gamma_\nu \gamma_5), \qquad G_5 (1 \otimes 1) .
\label{vec II}
\end{equation}

Turning to the isovector axial current, the type-I terms are again present.
These involve the matrix combinations:
\begin{eqnarray}
G_1 \epsilon^{abc} (\tau^c \otimes i\gamma_5 \tau^b), &&\qquad G_2 (\gamma^\nu
\gamma_5 \otimes \gamma_\nu \tau^a + \gamma^\nu \otimes \gamma_\nu \gamma_5
\tau^a), \qquad G_3 (\gamma^\nu \gamma_5 \tau^a \otimes \gamma_\nu),
\nonumber\\
G_4 (\gamma^\nu \tau^a \otimes \gamma_\nu \gamma_5),&&\qquad G_5 
\epsilon^{abc} (i\gamma_5\tau^b \otimes \tau^c) . \label{ax I}
\end{eqnarray}
There are no type-II pieces in this current, but a third kind of nonlocal
structure does occur,
\begin{eqnarray}
J^\mu_{(III)}&=&\frac{i}{(2\pi)^{12}} \sum_i G_i \int \prod_n d^4
p_n ~\, \overline{\psi} (p_1) \Gamma_i^\alpha \psi (p_3) \overline{\psi} (p_2) 
\Omega_{i\alpha} \psi (p_4) ~\, \delta(p_1 + p_2 + q - p_3 - p_4)\nonumber\\
&&\qquad\qquad\times\int_0^1 d\lambda \left[ f(p_2)f(p_3)\frac{\partial}
{\partial (p_1 + p_4)_\mu} f(p_1 + q - \lambda q) f(p_4 - \lambda q)  \right. 
\nonumber \\
&&\qquad\qquad\qquad\qquad\left. -f(p_1)f(p_4)\frac{\partial}{\partial 
(p_2 + p_3)_\mu}f(p_2 + \lambda q)f(p_3 - q + \lambda q) \right] . \label{III}
\end{eqnarray}
The relevant terms in this case are:
\begin{equation}
G_1 (i\gamma_5 \tau^a \otimes 1),\qquad G_2 \epsilon^{abc} (\gamma^\nu 
\gamma_5 \tau^c \otimes \gamma_\nu \tau^b),\qquad G_5 (i\gamma_5 \otimes 
\tau^a). \label{ax III}
\end{equation}

It is straightforward to see that dependence on the path, $\lambda$, does not
appear in the longitudinal components of the currents. Since the form factor
depends only on the square of its argument, one has in the case of type-I
contributions
\begin{equation}
q_\mu \frac{\partial}{\partial (p_1 + p_3)_\mu} f(p_1 + \lambda q) f(p_3 - q + 
\lambda q) = \frac{1}{2} \frac{d}{d\lambda} f(p_1 + \lambda q) f(p_3 - q + 
\lambda q) .  \label{long deriv}
\end{equation}
The $\lambda$ integral in $q_\mu J^\mu_{(I)}$ is then trivial, and produces a 
difference in form factors. Similar results can be seen to hold for the 
longitudinal components of the other nonlocal structures~(\ref{II},\ref{III}). 

Useful checks on the above expressions for the currents are provided by
various Ward identities which follow from (partial) current
conservation\cite{Pla}. In the case of the axial current, an extension of the
arguments in Ref.\cite{BB95} can be used to show that the
Gell-Mann--Oakes--Renner (GMOR) relation\cite{GMOR} holds. For the vector 
currents, we have checked that the two-point correlator of vector currents is 
purely transverse, that the $\gamma qq$ Ward identity is satisfied, that the 
pion charge is unity, and that the low-energy theorem for the anomalous decay
$\pi^0 \to \gamma\gamma$ is satisfied.

\section{Propagators and couplings to currents}\label{s3}
\subsection{Quark propagator}\label{quark}

The first step in the calculations is to construct the dressed quark 
propagator, by means of the corresponding Schwinger--Dyson equation (SDE). We
treat it in the ladder approximation, which is equivalent to working at leading
order in the $1/N_c$ expansion. Fig.~1 gives an illustration of the diagrams 
that are summed in this approximation. In terms of a momentum-dependent quark 
``mass" $m(p)$ defined from the dressed quark propagator by
\begin{equation}
S_F^{-1} (p)=\ps - m(p) , \label{q prop}
\end{equation}                 
this equation can be written as\footnote{Here `Tr' is used to denote a trace 
over flavour, colour and Dirac indices; the symbol `tr' will later be used to
indicate a trace over Dirac indices only.}
\begin{equation}
m(p)=m_c + i G_1 f^2(p) \hbox{Tr} \int \frac{d^4 k}{(2\pi)^4} \frac{\ks + 
m(k)}{k^2 - m^2(k)}f^2(k) . \label{SDE ladder}
\end{equation}
The dressing at this order occurs only through the interaction in the isoscalar
scalar channel, $G_1$. The integral here is very similar to that appearing in 
the quark condensate, apart from the extra interaction form factors. Just as in
the original NJL model, the dynamical generation of quark masses is intimately 
connected with the appearance of a non-zero condensate. In our numerical 
treatment of the model we evaluate loop integrals like that in (\ref{SDE 
ladder}) in Euclidean space, since the form factor has been defined for 
Euclidean momenta. Physical results are then obtained by analytically 
continuing back to Minkowski space. Notice that the separable nature of the 
interaction produces a great simplification since the dependence on the 
external momentum $p$ factorizes out of the loop integral.

The solution to the SDE (\ref{SDE ladder}) can be written simply as
\begin{equation}
m(p)=m_c + \Bigl(m(0) - m_c\Bigr)f^2(p) .
\end{equation}
Hence to obtain the quark mass it is necessary to determine only the constant
$m(0)$. This can be done easily using iterative methods. In practice we choose 
to use equation~(\ref{SDE ladder}) to determine the parameter $G_1$ for a given
value of $m_0(0)$, the zero-momentum quark mass in the chiral limit.

The denominator of the quark propagator, $p^2 - m^2(p^2)$, does not have a
zero at positive (Minkowski) $p^2$ if $m(0)$ is sufficiently large. This
property provides a sufficient, although not strictly a necessary, condition
for confinement. There are still poles in the quark propagator, but they are
shifted into the complex $p^2$ plane. Such behaviour is not uncommon in models
of quark confinement based on the solution of a Schwinger--Dyson
equation~\cite{FR96,SDErev,Sti} in the ladder approximation. Because of the
simplifications due to the separable interaction, the present model provides a
convenient setting in which to investigate this mechanism for confinement. As
pointed out by Lee and Wick~\cite{LW69} (see also\cite{CLOP}), particles which
have a complex mass of this type should not appear as asymptotic states if one
is to have a unitary S-matrix. When amplitudes have been defined in Euclidean
space, the prescription for analytically continuing them back to Minkowski
space must respect this requirement, as described in more detail in
Subsec.~\ref{loopint}.

\subsection{Meson propagators}\label{meson props}
The meson masses and vertex functions are found using the Bethe--Salpeter
equation (BSE). For consistency with the SDE (\ref{SDE ladder}), we work in
the ladder approximation. At this level, the separable interaction allows the
$\overline{q}q$ scattering matrix, $T$, to be written in the form
\begin{equation}
T(p_1,p_2,p_3,p_4) = \prod_n f(p_n) ~\, \delta(p_1 + p_2 - p_3 - p_4) 
\hat T(q),
\end{equation}
where we have denoted the total momentum of the $\overline{q}q$ pair by $q=p_1
- p_3=p_4 - p_2$. The BSE, shown schematically in Fig.~2, may be conveniently
expressed in terms of $\hat T$ as
\begin{equation}
\hat T(q)=G + GJ(q)\hat T(q) , \label{BSE matrix}
\end{equation}
where $G$ is simply a matrix of the coupling constants from the action
(\ref{4q action}--\ref{separable}) and $J(q)$ is composed of the loop
integrals
\begin{equation}
J_{ij}(q)=i \hbox{Tr} \int \frac{d^4 p}{(2\pi)^4} f^2(p_+) f^2(p_-) \Gamma_i 
S_F(p_-) \Gamma_j S_F(p_+),  \label{J}
\end{equation}
where we have introduced $p_\pm=p\pm {1\over 2}q$. The quark propagators to be
used in~(\ref{J}) are the dressed propagators obtained by solving the ladder
SDE.

The mesonic bound states are located at the poles of $\hat T$. These can be 
determined from the equation
\begin{equation}
\hbox{det} (1 - GJ(q))=0 . \label{boundstates}
\end{equation}                                                      

The form of the action means that the matrix $G$ is diagonal with respect to
flavour and Lorentz structures. The scattering matrix, however, is only
block-diagonal, since certain off-diagonal elements of $J$ may be non-zero.
In particular there is a loop integral that leads to mixing between the
pseudoscalar and longitudinal axial channels. This $\pi a_1$ (and $\eta^\star
f_1$) mixing is an example of the ``partial Higgs mechanism" familiar from
effective Lagrangians of $\pi, \rho$ and $a_1$ mesons\cite{Mei88,Bir96}. It
leads to an axial as well as a pseudoscalar vertex function for the pion and
$\eta^\star$. With each flavour having an equal current quark mass, no such 
mixing occurs between the scalar and vector channels, as can be seen from the 
fact that the integrand in the corresponding element of (\ref{J}) is odd under 
$p \to -p$. The absence of such a mixing means that the longitudinal vector 
channel is quite independent of the scalar one. It is therefore important to 
check that a pole does not develop in the former channel, since that would be 
unphysical.

For later ease of reference, we label the various elements of $J$ as
follows for the Dirac matrices inserted:
\begin{eqnarray}
J_{SS}:\; 1 \otimes 1,\qquad J_{VV}^{T}:&&\; T_{\mu\nu}(\gamma^\mu \otimes 
\gamma^\nu),\qquad J_{VV}^{L}:\; q^{-2} (-i\qs \otimes i\qs),\nonumber\\
J_{PP}:\; i\gamma_5 \otimes i\gamma_5,\qquad J_{AP}:&&\; m_\pi^{-1} 
(-i\qs\gamma_5 \otimes i\gamma_5),\qquad J_{PA}:\; m_\pi^{-1}(i\gamma_5 
\otimes i\qs \gamma_5), \nonumber\\
J_{AA}^T:\; T_{\mu\nu}(\gamma^\mu \gamma_5 \otimes \gamma^\nu \gamma_5),
&&\qquad J_{AA}^{L}:\; m_\pi^{-2} (-i\qs\gamma_5 \otimes i\qs \gamma_5) ,
\end{eqnarray}
where $T_{\mu\nu}$ is the transverse projector,
\begin{equation} 
T_{\mu\nu}=g_{\mu\nu} - \frac{q_\mu q_\nu}{q^2}. \label{tmunu}
\end{equation}
In this basis, $J_{AP}=J_{PA}$.

Near to the pole corresponding to a particular particle the relevant part of
$\hat T$ may be represented as
\begin{equation}
\frac{\overline{V}(q) \otimes V(q)}{m^2 - q^2}, \label{vertex fn}
\end{equation}
where $V(q)$ denotes the appropriate vertex function and any polarization
indices have been suppressed. The vertex functions for the final and initial
states are related by $\overline{V}= \gamma^0 V^\dagger \gamma^0$.
For the particles of interest, they are:
\begin{eqnarray}
V_\pi (q)=(g_{\pi qq} - m_\pi^{-1} \widetilde{g}_{\pi qq} \qs)i\gamma_5 
\tau^a,&&\qquad V_\sigma (q)= g_{\sigma qq},\nonumber\\
V_{\rho s}(q)= g_{\rho qq} \epsilons_s \tau^a,\qquad V_{a_1 s}(q)= g_{a_1 
qq} \epsilons_s \gamma_5 \tau^a,&&\qquad V_{\omega s} (q)=g_{\omega qq} 
\epsilons_s, \nonumber\\
V_{\eta^\star} (q)=(g_{\eta^\star qq} - m_{\eta^\star}^{-1} 
\widetilde{g}_{\eta^\star qq} \qs )i\gamma_5,&&\qquad V_{a_0} (q)= g_{a_0 qq} 
\tau^a . 
\end{eqnarray}
For all particles except the pseudoscalars there is no mixing, and so each has
a single coupling constant $g_{iqq}$ describing the coupling of an on-shell
meson to quarks. These couplings are related to the corresponding loop
integrals (\ref{J}) by
\begin{equation}
{1\over g_{iqq}^2}=\left. (-1)^{S} \frac{dJ_{ii}}{dq^2} \right|_{q^2=m^2}, 
\label{meson couplings}
\end{equation}
where $S$ is the spin of the meson. The couplings of the pion to quarks,
$g_{\pi qq}$ and $\widetilde{g}_{\pi qq}$, are given by:
\begin{equation}
g^2_{\pi qq}=-G_1 \frac{\Bigl(1 - G_2 J_{AA}^{L}(m^2)\Bigr)}
{D^\prime_\pi (m^2)},\qquad
g_{\pi qq} \widetilde{g}_{\pi qq}=\frac{G_1G_2J_{PA}(m^2)}
{D^\prime_\pi (m^2)},
\label{pi couplings}
\end{equation}
where the prime indicates a derivative with respect to $q^2$ and the
pseudoscalar-axial determinant $D_\pi(q^2)$ is defined to be:
\begin{equation}
D_\pi (q^2)=\Bigl(1 - G_1J_{PP}(q^2)\Bigr)\Bigl(1 - G_2 J_{AA}^{L}(q^2)\Bigr) 
- G_1G_2 J_{AP}^2(q^2) . \label{piondet}
\end{equation}
Similar expressions hold for the couplings of the $\eta^\star$, with $G_5$ and
$G_4$ playing the roles of $G_1$ and $G_2$ respectively.

\subsection{Loop integrals}\label{loopint}
In Euclidean space, the loop integrals in the BSE (\ref{J}) take the form
\begin{equation}
J_{ij}(q^2)=-N_cN_f \int \frac{d^4 p}{(2\pi)^4} \frac{f^2(p_+) f^2(p_-)
t_{ij}(p^2,q^2,p\cdot q)}{(p_+^2 + m_+^2)(p_-^2 + m_-^2)} ,
\label{generic loopint}
\end{equation}
where $t_{ij}$ is the appropriate Dirac trace and all momenta are to be 
understood as Euclidean. We have introduced $m_\pm$ here to denote the quark
mass evaluated at $p_\pm$. Consider this integral evaluated at some timelike
momentum, $q=(\ZEROVEC,iq_0)$. For a confining parameter set, each quark
propagator considered as a function of energy has four poles at complex
energies corresponding to a pair of complex-conjugate poles in $p^2$. As $q_0$ 
is increased these poles are translated parallel to the imaginary $p_4$ axis. 
For any given value of $|\underline{p}|$, there is value of $q_0$ for which the
poles of the $p_-$ and $p_+$ quark propagators meet on the real $p_4$ axis, 
pinching the contour of integration. For larger values of $q_0$ the poles cross
the axis and may contribute an imaginary part to the propagator in the meson 
channel, depending on the prescription used to continue the integral beyond the
pinch point.

The usual prescription based on Wick rotation of the integration contour would 
lead to an imaginary part of the meson propagator starting at the minimum value
of $q_0$ for which the contour is pinched. This would correspond to the
opening of a threshold for decay of a meson into other states, and as described 
above (Subsec.~\ref{quark}) is inappropriate here.

One possible continuation, originally suggested by Cutkosky et al.~\cite{CLOP},
amounts to the deformation of the integration contour displayed in Fig.~3.
This prescription ensures that the resulting meson propagator does not develop
an imaginary part above the pseudo-threshold where the contour is pinched by
the complex poles, although it does mean that the propagator cannot be
analytically continued past that point. Since the method is not unique, the
choice of continuation prescription must be regarded as an additional
assumption that forms part of the specification of any model with a quark
propagator of this type. In our calculations we shall adopt the suggestion of
Cutkosky et al., which was shown in Ref.~\cite{CLOP} to be consistent with the
requirements of unitarity and macrocausality.

As discussed by both Cutkosky et al.~\cite{CLOP} and Lee and Wick~\cite{LW69},
microcausality violations can occur in models with a Euclidean metric and
states of complex mass. However, in order to measure such violations, Lee and
Wick~\cite{LW69} have estimated that one would need to create a wave packet of
width $\ll \gamma^{-1}$, where the complex mass is $M + \frac{1}{2} i \gamma$.
In any event, microcausality in the model is intrinsically broken by the use
of an action with nonlocal interactions.

In evaluating these integrals numerically, we take a contour in $p_4$ that
runs along the real axis. For energies $q_0$ above the pseudo-threshold, the
prescription we use means that we have to add in contributions from the
residues of the poles that have crossed the axis. Both the naive integral over
Euclidean four-momentum in (\ref{generic loopint}) and the residue
contributions diverge at the pinch point, although these divergences cancel to
leave a finite result\cite{CLOP}. The cancellation occurs at the level of the
integrated result rather than at all values of the three-momentum flowing
round the loop. In our numerical work, we therefore need to regulate the two
contributions when evaluating them separately. We subtract off a function which
has the same divergence as the naive integral when $p$ lies within a radius 
$\Delta$ of the pinch point. A similar function is used to cancel the divergent
part of the residue contribution when $|\underline{p}|$ is less than $\Delta$
from its pinch value and is chosen to cancel exactly with the piece that has
been cut out of the naive integral. We have checked that our results are
independent of the value of the regularizing parameter $\Delta$.

It should be noted that the quark propagator of the present model has many
complex-conjugate pairs of poles. Such an analytic structure is also found in
the pion propagator of the NJL model within the proper-time regularization
scheme\cite{BRNG96}. In our model these poles are present for both confining
and non-confining parameter sets and their positions depend on the detailed
behaviour of the form factor for large momenta. We therefore regard them as
unphysical artifacts of the model. For the parameter sets considered here,
the next set of poles would result in another pseudo-threshold at energies
of $\sim 2$ GeV. Since the model is not intended to be credible at such 
momenta, these extra poles do not present a practical problem. 

\subsection{Current couplings}\label{curr coupl}

The electromagnetic or weak decay constant of a meson is given by the matrix
element between the vacuum and that meson of the vector or axial current. In
the present model one needs to include contributions from both the usual local
current and the nonlocal pieces described in Sec.~\ref{s2}. The corresponding
diagrams are shown in Fig.~4.

We consider first the pion decay constant. The loop integral arising from 
the local part of the axial current is very similar to $J_{AP}$, except that 
only two (rather than four) form factors are present. One must also include a
contribution from the $G_1 (i\gamma_5 \tau^a \otimes 1)$ term with type-III
structure~(\ref{III}) in the axial current. As described in Ref.~\cite{BB95},
this piece can be written as a sum of terms each of which factorizes into a
loop integral somewhat similar to that in the scalar quark condensate, together
with a loop with a pseudoscalar insertion and a pion vertex function. It is
convenient to refer to these loops as being one-quark or two-quark, according
to the number of quark propagators involved. The contribution of the diagram to
$f_\pi$ is
\begin{eqnarray}
\frac{i G_1}{2m_\pi^2} \int \frac{d^4 k}{(2\pi)^4} \frac{\hbox{Tr}
[\ks + m(k^2)]}{k^2 - m^2(k^2)} \int \frac{d^4 p}{(2\pi)^4} 
\frac{\hbox{Tr} V_\pi(q) (\ps_- + m_-) 
\gamma_5 \tau^a (\ps_+ + m_+)}{(p_+^2 - m_+^2)(p_-^2 - m_-^2)} f(p_+) f(p_-)
\nonumber\\
\times\Bigl[f^2(k)\Bigl(f^2(p_+) + f^2(p_-)\Bigr) - f(p_+) f(p_-) f(k) 
\Bigl( f(k+q) + f(k-q)\Bigr)\Bigr].
\label{fpi scalar}
\end{eqnarray}

In the extended model there is another nonlocal contribution which is induced
by the term $G_2 (\gamma^\nu \otimes \gamma_\nu \gamma_5 \tau^a)$ with type-I
structure~(\ref{I}) in the axial current. The one-quark loop in this case has
a vector insertion. Although the vacuum expectation value of $\overline{\psi}
\gamma^\nu \psi$ vanishes by Lorentz invariance, a non-zero integral is
produced by a combination of form factors that is anti-symmetric in the loop
momentum. The contribution of this diagram to $f_\pi$ is
\begin{eqnarray}
\frac{-i G_2}{2m_\pi^2} &&\int \frac{d^4 k}{(2\pi)^4} \frac{\hbox{Tr} 
\gamma^\nu [\ks + m(k^2)]}{k^2 - m^2(k^2)} f(k) \Bigl(f(k+q) - f(k-q)\Bigr)
\nonumber\\
\times&&\int \frac{d^4 p}{(2\pi)^4} \frac{\hbox{Tr} V_\pi(q) (\ps_- + m_-) 
\gamma_\nu \gamma_5 \tau^a (\ps_+ + m_+)}
{(p_+^2 - m_+^2)(p_-^2 - m_-^2)} f^2(p_+) f^2(p_-) . \label{fpi vector}
\end{eqnarray}
These pieces of $f_\pi$ arising from the nonlocal current make significant
numerical contributions and are needed in order to satisfy the 
Gell-Mann--Oakes--Renner relation\cite{BB95,Pla}.

A determination of the coupling strength of the $a_1$ particle to the 
transverse axial current requires the evaluation of very similar diagrams to
those involved in $f_\pi$. However, an important difference from the analogous 
nonlocal diagrams for $f_\pi$ arises because the path-dependent integral over 
$\lambda$ for the transverse current is non-trivial. In this case a numerical 
integration over $\lambda$ is also required. With a type-I nonlocal structure 
(\ref{I}), $\lambda$ is only present in form factors associated with one of the
loops. This is not so for contributions induced by type-II (\ref{II}) or
type-III (\ref{III}) terms in the current, where the integrals for the one-
and two-quark loops do not factorize.
 
The couplings responsible for the electromagnetic decays of the vector mesons
can be evaluated in a similar manner. Again the nonlocal contributions are
numerically important and are essential if Ward identities such as
\begin{equation}
q^\mu i\int d^4 x e^{iqx} \langle 0|T \{ J_\mu^a(x) {\cal O}(0) \} |0 \rangle
=0
\end{equation}
are to be satisfied. 

A common scheme to describe electromagnetic couplings in the 
literature~\cite{BCexamples,KK96,FMRT95}, sometimes called the impulse approximation, 
involves the $\gamma qq$ vertex only, neglecting irreducible
couplings of the photon to more than two quarks. The $\gamma qq$ vertex itself
is chosen to be of the Ball--Chiu~\cite{BC80} form. For the model studied
here, use of the impulse approximation is not an appropriate prescription for
the calculation of electromagnetic observables. For example, as 
discussed in Sec.~\ref{s6}, it would not yield the correct value for the pion
charge. In this model, the electromagnetic couplings are specified once a
particular ansatz has been chosen for the nonlocal part of the vector current.
The uncertainty inherent in the construction of the transverse part of the
current is discussed in Sec.~\ref{s2}. Despite this, gross features of the
nonlocal current would remain unchanged with different path ansatze.

The dressed $\gamma qq$ vertex in this model is described in some
detail below since it is a necessary ingredient in the calculation of many
electromagnetic processes. An important constraint on the structure of this
vertex, $\Gamma_\mu$, is provided by the Ward-Takahashi identity
\begin{equation}
q^\mu \Gamma_\mu(p,q)=S_F^{-1}(p_+) - S_F^{-1}(p_-) , \label{gammaqq WI}
\end{equation}
where $q$ is the photon momentum flowing away from the vertex and $p$ is the
momentum flowing through the vertex. (The isospin structure has been suppressed
here.)

The various pieces of the full $\gamma qq$ vertex within this
approach are shown diagrammatically in Fig.~5. From the local current, there
is simply a contribution to $\Gamma_\mu$ of the usual form, $\gamma_\mu$. The
nonlocal current induces contributions where there is a closed one-quark loop,
similar to those in the pion decay constant discussed above. In the
electromagnetic case, the diagram where the closed loop has a scalar insertion
can be simplified by using the SDE~(\ref{SDE ladder}) to express it in the form
\begin{equation}
- \Bigl(m(0) - m_c\Bigr) \int_0^1 d\lambda \frac{\partial}{\partial p^\mu} 
f^2 \Bigl(p +(\lambda - {\textstyle\frac{1}{2}}) q\Bigr) . \label{gammaqq sc}
\end{equation}
Added to the local contribution, this would give the full vertex in a version
of the model without vector mesons. Now, as a special case of Eq.~(\ref{long
deriv}), we have that
\begin{equation}
q^\mu \frac{\partial}{\partial p^\mu} f^2\Bigl(p + (\lambda - 
{\textstyle\frac{1}{2}}) q\Bigr) = \frac{d}{d\lambda} f^2\Bigl(p + (\lambda - 
{\textstyle\frac{1}{2}}) q\Bigr) .
\end{equation}
The $\lambda$ integral involved in $q^\mu \Gamma_\mu$ is therefore trivial, and
so the Ward identity for the $\gamma qq$ vertex (\ref{gammaqq WI})
can be seen to be satisfied by the sum of $\gamma^\mu$ and (\ref{gammaqq sc}).

In the extended model with vector-meson degrees of freedom, there is another 
contribution to $\Gamma_\mu$ involving a one-quark loop with a vector
insertion, specifically,
\begin{equation}
-i \gamma_\nu f(p_-)f(p_+) G_2 N_c N_f \int \frac{d^4 k}{(2\pi)^4} 
\frac{4k^\nu}{k^2 - m^2(k^2)} \int_0^1 d\lambda \frac{\partial}{\partial k^\mu}
f(k - q + \lambda q)f(k + \lambda q) . \label{gammaqq vec}
\end{equation}
In addition there are pieces which contain the propagator of an intermediate
$\overline{q}q$ state, as constructed from the ladder BSE~(\ref{BSE matrix}),
in the vector channels. The contribution of the longitudinal channel to 
$\Gamma_\mu$ is
\begin{eqnarray}
i \frac{\qs}{q^2} f(p_-)f(p_+) \frac{G_2 N_c N_f}{1 - G_2 J_{VV}^L (q)} \int
&&\frac{d^4 k}{(2\pi)^4} \frac{f(k_-)f(k_+)}{(k_-^2 - m_-^2)(k_+^2 - m_+^2)}
\nonumber\\
&&\times\hbox{tr} \widetilde{\Gamma}_\mu (k,q) (\ks_- + m_-) \qs (\ks_+ + m_+),
\label{gammaqq vl}
\end{eqnarray}
while the transverse channel gives
\begin{eqnarray}
i \left(\gamma^\nu - \frac{q^\nu \qs}{q^2} \right) f(p_-)f(p_+) \frac{G_2 
N_c N_f}{1 - G_2 J_{VV}^T (q)} \int &&\frac{d^4 k}{(2\pi)^4} 
\frac{f(k_-)f(k_+)}{(k_-^2 - m_-^2)(k_+^2 - m_+^2)}\nonumber\\
&&\times\hbox{tr} \widetilde{\Gamma}_\mu (k,q) (\ks_- + m_-) \gamma_\nu 
(\ks_+ + m_+) ,\label{gammaqq vt}
\end{eqnarray}
where $\widetilde{\Gamma}^\mu (k,q)$ is the two-quark-irreducible $\gamma qq$
vertex consisting of the sum of $\gamma^\mu$, (\ref{gammaqq sc})
and~(\ref{gammaqq vec}). In these expressions, $m_\pm$ denotes the quark
mass evaluated at $k_\pm$.

To check that the additional contributions (\ref{gammaqq vec}-\ref{gammaqq
vt}) in the extended model are consistent with the Ward identity~(\ref{gammaqq
WI}), we note first that the quark propagator is unchanged and so the sum of
the local piece and (\ref{gammaqq sc}) still saturates the identity. In the
contribution of expression~(\ref{gammaqq vec}) to $q^\mu \Gamma_\mu$,
Eq.~(\ref{long deriv}) enables the integration over the path variable to be
performed. This part of $q^\mu \Gamma_\mu$ is then
\begin{equation}
-i \frac{\qs}{q^2} f(p_-)f(p_+) G_2 N_c N_f \int \frac{d^4k}{(2\pi)^4} 
\frac{4q\cdot k}{k^2 - m^2(k^2)} f(k) \Bigl(f(k+q) - f(k-q)\Bigr) . 
\label{WI cancel}
\end{equation}                  
The purely transverse piece (\ref{gammaqq vt}), which involves a
propagating $\rho$-meson, obviously does not contribute to the Ward identity.
Thus cancellation must occur between (\ref{WI cancel}) and the piece coming
from (\ref{gammaqq vl}). To demonstrate this explicitly, we make use of the
fact that $q^\mu \widetilde{\Gamma}_\mu$ is given by the sum of (\ref{WI
cancel}) and the right-hand-side of (\ref{gammaqq WI}). The contribution to 
$q^\mu \Gamma_\mu$ involving the longitudinal $\overline{q}q$ intermediate 
state (\ref{gammaqq vl}) can then be expressed as
\begin{eqnarray}
i \frac{\qs}{q^2}\frac{G_2N_cN_f}{1 - G_2J_{VV}^L (q)}&& f(p_-)f(p_+)\left[  
\int \frac{d^4k}{(2\pi)^4} \frac{\hbox{tr} (\qs + m_- - m_+)(\ks_- + m_-)
\qs(\ks_+ + m_+)}{(k_+^2 - m_+^2)(k_-^2 - m_-^2)} f(k_+)f(k_-)\right.
\nonumber\\
&&\left. -G_2J_{VV}^L (q) \int \frac{d^4k}{(2\pi)^4} \frac{4q\cdot k}
{k^2 - m^2(k^2)} f(k) \Bigl(f(k+q) - f(k-q)\Bigr)\right] . 
\label{WI cancel?}
\end{eqnarray}
The Dirac trace in the first line of the above expression may be 
written as
\begin{equation}
4(q\cdot k_-) (k_+^2 - m_+^2) - 4(q\cdot k_+) (k_-^2 - m_-^2) .
\end{equation}
Hence in each of the resulting terms, one of the factors $k_\pm^2-m_\pm^2$ can
be cancelled in the denominator of the integral. Shifting the integration 
variable from $k$ to $k_\pm$ as appropriate, the first term inside the square
brackets of~(\ref{WI cancel?}) may be rewritten in the same form as the
second, demonstrating the required cancellation. Note that the discussion
above has referred to the presence of the $G_2$ coupling in the isovector
interaction channel. The results in the isoscalar channel are completely
analogous, with the replacement of $G_2$ by $G_3$.

For the purpose of practical calculations, it is convenient to collect 
together the various contributions to the vertex in the form
\begin{eqnarray}
\Gamma_\mu (p,q) &=& \gamma_\mu Q + \left( \gamma_\mu - \frac{q_\mu \qs}{q^2} 
\right) f(p_-)f(p_+)B(q^2) \nonumber\\
&&\qquad\qquad- 2 Q \int_0^1 d\lambda (p + (\lambda - 
{\textstyle\frac{1}{2}}) q)_\mu m^\prime (p + (\lambda - 
{\textstyle\frac{1}{2}}) q) , \label{gammaqq full}
\end{eqnarray}
where the prime denotes a derivative with respect to the square of the momentum
argument and we have reinstated the flavour structure using the charge matrix 
$Q= {1 \over 2} (\tau^3 + {1 \over 3})$. The function $B(q^2)$ accounts for the
presence of vector mesons in the model and has the form $B(q^2)= {1 \over 2} 
(\tau^3 B_2(q^2) + {1 \over 3} B_3(q^2))$ where the functions $B_i(q^2)$ are
\begin{eqnarray}
B_i(q^2) &=& \left\{ \frac{1}{1 - G_i J_{VV}^T (q)} \right\}
\left\{ A_i(q) + i G_i N_c N_f \int \frac{d^4 k}{(2\pi)^4} 
\frac{f(k_-)f(k_+)}{(k_-^2 - m_-^2)(k_+^2 - m_+^2)} \right.\nonumber\\
&&\qquad\qquad\times\left[ \left( 4m_-m_+ - 4k^2 + q^2 + \frac{8}{3} \left( k^2
- \frac{(q\cdot k)^2}{q^2} \right) \right) \right.\nonumber\\
&&\qquad\qquad\qquad\left. \left. - \frac{8}{3}(m_+ + m_-) \left( k^2 - 
\frac{(q\cdot k)^2}{q^2} \right)\int_0^1 d\lambda\; m^\prime \Bigl(k + (\lambda 
- {\textstyle\frac{1}{2}}) q\Bigr) \right] 
\right\}  , \label{b2}
\end{eqnarray}
and the $A_i(q^2)$ in the above equation originate from the one-quark loop with
a vector insertion and are given by
\begin{eqnarray}
A_i(q^2)&=& - \frac{8i}{3} G_i N_c N_f \int \frac{d^4 k}{(2\pi)^4} 
\frac{k^2 - (q\cdot k)^2/q^2}{k^2 - m^2(k^2)}\nonumber\\
&&\qquad\times\int_0^1 d\lambda \Bigl(f^\prime(k + \lambda q)f(k - q + \lambda 
q)+  f(k + \lambda q)f^\prime(k - q + \lambda q)\Bigr) . \label{a2}        
\end{eqnarray}

Writing $A_i(q^2)$ and $B_i(q^2)$ in Euclidean-space form and
performing an integration by parts on (\ref{a2}), one finds that $B_i(0)=0$.
This is simply a consequence of the differential form of
(\ref{gammaqq WI}),
\begin{equation}
\Gamma_\mu (p,0) = Q\frac{\partial}{\partial p^\mu} S_F^{-1}(p) .
\end{equation}
Hence, in evaluating processes where the photon is on-shell, the $\gamma qq$ 
vertex is unchanged by the existence of vector-meson degrees of freedom in the
model. 

For the coupling of an on-shell vector meson to the vector current, one
evaluates a quark loop between the meson vertex function and that part of the
$\gamma qq$ vertex which does not include the contribution from the
propagating transverse vector channel,~(\ref{gammaqq vt}). Attempting to
include such a diagram would merely amount to the addition of another bubble
onto the vector-meson chain. Since (\ref{gammaqq vt}) is purely transverse, the
Ward identity for the vertex still holds.

\section{Numerical fits}\label{s4}

The model as defined in Sec.~\ref{s2} above has seven parameters: the
current quark mass ($m_c$), the range of the form factor ($\Lambda$) and five
possible coupling constants. Considering first the couplings $G_1$ and $G_2$
only, we have chosen to fit the quantities: $m_\pi=140$ MeV, $f_\pi=93$ MeV
and $m_\rho=770$ MeV, which do not depend on the remaining three couplings.
This leaves one parameter undetermined which we have taken to be $m_0(0)$, the
zero-momentum quark mass obtained in the chiral limit of the SDE (\ref{SDE
ladder}). Each of the other three couplings may then be independently fixed to
reproduce the mass of the corresponding meson: $G_3$ is set by requiring
$m_\omega=783$ MeV ; $G_4$ by $m_{f_1}=1282$ MeV and $G_5$ by $m_{a_0}=982$
MeV. The meson masses are determined from Eq.~(\ref{boundstates}), whilst
$f_\pi$ is given by the coupling of the pion to the axial current. The
contributions to $f_\pi$ from the nonlocal pieces of the current are 
significant: the scalar and vector loop terms described in the previous section
are respectively $\sim 35 \%$ and $\sim -10 \%$ of the total value. 

In terms of $m_0(0)$, the possible fits have a restricted range. Having
a coupling strong enough to realize confinement requires that $m_0(0) \gapprox 
270$ MeV. Below that value, the model should only be used up to an energy
corresponding to the appearance of the $\overline{q}q$ continuum at twice the 
value of the (purely real) quark mass. In fact only a very limited range of 
non-confining sets are possible because the empirical masses of the vector 
mesons are located in this continuum for $m_0(0) \lapprox 250$ MeV. 

An upper limit on the acceptable values for $m_0(0)$ is imposed by the
behaviour of the meson propagators above the pseudo-threshold. The dramatic
changes of behaviour which can occur beyond this point may be seen in
Fig.~6, where the denominators of the propagators in various scattering 
channels are plotted for a parameter set with $m_0(0)=300$ MeV. In this case 
the pseudo-threshold occurs at an energy of 895 MeV. For larger values of 
$m_0(0)$ this energy decreases. As suggested by the behaviour in Fig.~6, for 
large enough $m_0(0)$ two additional poles appear in the transverse-vector 
channel above the $\rho$ pole. The first of these extra poles has a residue of 
the wrong sign to describe a physical particle. Although one might hope to
consider parameter sets with the extra poles, provided that they lie well above
the energies of interest, in practice this is possible only for values of
$m_0(0)$ within a very narrow range, $\sim 320-330$ MeV.

A pronounced change in behaviour beyond the pseudo-threshold also occurs in
the longitudinal-vector channel and seems to be important in ensuring that no
poles are present in this channel. An unphysical pole does occur, however, in
the pion propagator (Fig.~6). This unwanted pole is located between $1.3$ and
$1.45$ GeV, depending on the parameter set used. One should therefore only
attempt to use the model at smaller energies. Although a similar pole also
appears in the $\eta^\star$ propagator, it lies at a higher energy than in the
pion case.
	
We present here numerical results for parameter sets which lie near each edge
of the acceptable range for $m_0(0)$. The variation of results over the full
range is generally monotonic; where it is not, the dependence on $m_0(0)$ is
fairly weak. Specifically, we have chosen to work with $m_0(0)=280$ MeV
(henceforth referred to as set A) and $320$ MeV (set B). Details of these
parameter sets are given in Table~1. Values of the zero-momentum quark masses
calculated for these parameters at non-zero $m_c$ are also quoted in Table~1.

Including a non-zero $m_c$ in the SDE has a significant effect on the
solution, a current mass of $\sim 10$ MeV leading to an increase of $\sim 50$
MeV in the zero-momentum dynamical quark mass. It is thus worth examining the
deviations of $f_\pi$ and $m_\pi$ from the values which would be obtained at
leading order in the chiral expansion. If one sets $m_c$ to zero and keeps
all other parameters fixed, then the corresponding values for the chiral limit
of $f_\pi$ are 84.6 MeV for set A and 85.1 MeV for set B. To leading order
in $m_c$, the pion mass is given by the GMOR relation
\begin{equation}
m_\pi^2 f_\pi^2=-m_c \langle \overline{\psi} \psi \rangle_0,
\label{GMORrel}
\end{equation}  
where $f_\pi$ and the quark condensate are evaluated in the chiral limit. This
expression gives values for $m_\pi$ of 143.6 MeV for set A and 143.2 MeV for
set B. Comparing these results with those of the full calculations for these
parameter sets, we see that higher-order terms in the chiral expansion
contribute about 10\% to $f_\pi$. In contrast, the corresponding contributions
to $m_\pi$ are less than 3\%. These observations suggest that the restoring
forces against deviations from the chiral circle are rather weak in this
model. One would therefore expect to find a light $\sigma$ meson in connection
with this softness of the vacuum. This is indeed the case, as will be
discussed shortly.

In the chiral limit, the model quark condensate is $- (206 \hbox{MeV})^3$ and
$- (189 \hbox{MeV})^3$ for sets A and B respectively. With non-zero
current-quark mass, the condensate integral is quadratically divergent. If it
is regulated by subtracting the perturbative condensate, slightly higher
values of $- (212 \hbox{MeV})^3$ and $- (193 \hbox{MeV})^3$ are obtained.
These are similar in size to values for the condensate determined from QCD sum
rules\cite{SVZ,GL82}. However one should bear in mind the fact that the 
condensate in QCD is a renormalized and scale-dependent quantity and so one
ought to be careful about comparing it directly with the value obtained in a
model of this type.

Table~2 lists the positions of the first few sets of poles in the 
quark propagator. Since we only consider the first group of poles to have 
physical relevance, the model should be used only for energies of less than
twice the real part of the second set of poles. This limit is $2.3$ GeV and
$1.9$ GeV for the parameter sets A and B, and so is sufficiently far above
the upper limit imposed by the unphysical pole in the pion channel not to be of
practical concern.

In Table~3, the calculated meson masses are given, along with their on-shell
couplings to quarks, as defined in Eqs.~(\ref{meson couplings}--\ref{piondet}).
The scalar isoscalar state is rather light. For comparison, the mass of the 
corresponding particle in the NJL model\cite{NJLrevs,ENJLrev,ENJLboson} is
$m_\sigma^2=m_\pi^2 + 4m^2$, where $m$ is the mass of the constituent quark. 
Interestingly, the $\sigma$ mass in our model varies only slowly with the 
dynamical quark mass. 

The empirical determination of the mass of the corresponding scalar isoscalar
meson is still unclear because of the very strong coupling between that meson
and the two-pion channel. While some analyses find masses of the order of 1
GeV\cite{amp}, others result in much smaller values\cite{sigma}. The $\sigma$
masses obtained in this model, like those in the NJL model, are compatible
with the latter. However, before taking them too seriously, one should
remember that the strong coupling to two pions also means that $1/N_c$
corrections are likely to be important in this channel\cite{NBCRG,BP97}. It is
thus premature to draw conclusions about the ability of the model to describe
this channel until such corrections have been calculated.

The calculated $a_1$ mass is somewhat smaller than the observed 1230
MeV\cite{PDG}. In the case of parameter set A, it lies a little below the
pseudo-threshold, but for most of the range of admissible $m_0(0)$ it is above
that energy. The $\rho-a_1$ mass splitting increases with increasing
constituent quark mass, although not so rapidly as suggested by the
NJL\cite{ENJLboson} expression, $m_{a_1}^2 = m_\rho^2 + 6m^2$, obtained from
the derivative expansion of the bosonized model. As a consequence of the
upper bound on the constituent mass, which follows from the effect of the
pseudo-threshold on the transverse-vector channel, it is not possible to
simultaneously reproduce the empirical values for both the $\rho$ and $a_1$
masses in the ladder approximation.

Since there are important flavour-mixing effects in the isoscalar pseudoscalar
sector, a realistic calculation for these mesons would require a three-flavour
version of the model. The $\eta^\star$ mass in the present two-flavour model
should not therefore be directly compared with experiment. It is nevertheless
reassuring that this mass lies between the physical $\eta$ and $\eta^\prime$
masses of $547$ and $958$ MeV respectively. Another possibly important feature
in the description of the state is the axial-pseudoscalar mixing with the
longitudinal $f_1$ channel. Indeed, in a Bethe-Salpeter study of a 
three-flavour model\cite{BQRTT}, the $\qs\gamma_5$ term in the vertex function 
was found to make significant contributions to the mass ($\sim 70$ MeV) and 
decay constant of the $\eta$. A similar effect has also been found in the NJL 
model\cite{KLVW}. In the present model, if the $f_1$ is omitted by setting 
$G_4$ to zero, then the $\eta^\star$ mass for parameter set A is reduced by 
around $20$ MeV, whereas for set B it falls by over $110$ MeV. These rather 
different behaviours are another consequence of the dramatic changes in the 
meson propagators which can occur at the pseudo-threshold. When $G_4=0$, the 
$\eta^\star$ mass lies below the pseudo-threshold for the full range of 
parameter sets. For non-zero $G_4$, the mixing increases the $\eta^\star$ mass 
and for parameter sets with $m_0(0)\gapprox 310$ MeV the mass is pushed above 
the pseudo-threshold, where the effect can be greatly enhanced. In addition,
for these parameter sets the gradient of the determinant $D_{\eta^\star}$
(cf.\ Eq.~(\ref{piondet})) changes significantly above the pseudo-threshold
and so the coupling of the $\eta^\star$ to quarks is considerably stronger.

\section{Hadronic decays}\label{s5}

At leading order in $1/N_c$, the three-meson vertices are calculated from
a quark loop with insertions of three vertex functions. We have obtained
several physical decay amplitudes from such diagrams with all of the mesons
on-shell, where the vertex functions are unambiguous. 

For an initial state of momentum $q$ decaying to particles with momenta $q_1$
and $q_2$, the quark propagators in the triangular loop are evaluated at $p
\pm \frac{1}{2}q$ and $p + \frac{1}{2} (q_1 - q_2)$. If the initial state has
a mass which is greater than twice the real part of the quark pole, then its
decay modes will be sensitive to pseudo-threshold effects. By analogy with the
loop integral in the BSE for that particle, residue contributions must be
taken into account in the three-point diagrams. It is also possible that
further residue contributions may be required if a final-state particle lies 
above the pseudo-threshold energy, but such a situation does not occur in 
practice for any of the amplitudes considered.

The couplings that we have considered are defined below, and their numerical
values are given in Table~4, along with the corresponding decay widths.
\begin{eqnarray}
\langle \pi^a(q_1) \pi^b(q_2) | \sigma(q) \rangle & = & - g_{\sigma\pi\pi}
\delta^{ab} \nonumber \\
\langle \pi^b(q_1) \pi^c(q_2) | \rho^a(q) \rangle & = & i g_{\rho\pi\pi}
\epsilon^{abc} (q_2\cdot \epsilon - q_1\cdot \epsilon) \nonumber \\
\langle \sigma(q_1) \pi^b(q_2) | a_1^a(q) \rangle & = & \frac{1}{2} i 
g_{a_1\sigma\pi} \delta^{ab} (q_1\cdot \epsilon - q_2\cdot \epsilon) \nonumber 
\\
\langle \rho^b(q_1) \pi^c(q_2) | a_1^a(q) \rangle & = & \epsilon^{abc} 
(g_{a_1\rho\pi} (\epsilon_\rho^\ast\cdot \epsilon_{a_1}) - h_{a_1\rho\pi}
(q_2\cdot \epsilon_\rho^\ast)(q_2\cdot \epsilon_{a_1})) \label{decays}
\end{eqnarray}

If the sigma meson of the model is to be interpreted as corresponding to the
scalar particle of the linear sigma model then its coupling to two pions
should be strong. Whilst the values in Table~4 do not indicate a particularly
broad state, this width is significantly reduced by the small available phase
space. To provide a useful comparison, we consider the prediction for the
two-pion coupling $g_{\sigma\pi\pi}$ from the linear sigma model. In that
model the coupling is $g_{\sigma\pi\pi}= (m_\sigma^2 - m_\pi^2) / f_\pi$
which, for the $\sigma$ masses of parameter sets A and B gives
$g_{\sigma\pi\pi}=1901$ MeV and $2122$ MeV respectively. These values are
$\sim 30 \%$ larger than the results for our model, indicating that the
coupling to pions of the scalar meson in the nonlocal NJL model is
qualitatively similar to that in the linear-sigma-model. As already 
mentioned, this strong coupling indicates the importance of going beyond 
leading order in $1/N_c$.
 
The $\rho$ meson decay width compares reasonably well with the empirical
value of $151$ MeV. In contrast, the corresponding leading-order
calculation in an extended NJL model, using the physical $\rho$ mass, 
significantly underestimates the decay rate\cite{rhodecay}. We find that it is 
not possible to simultaneously reproduce the empirical mass and decay width of
the $\rho$ at leading order in our model. However, if the model parameters for
a given $m_0(0)$ are refitted to the $\rho$ width rather than to its mass, 
the results for observables are not qualitatively different from those of
the original fit. For instance, this procedure would increase the $\rho$ mass
itself by $\sim 20$ to $60$ MeV.

The coupling $g_{a_1 \sigma\pi}$ is not a direct observable, although
the process it describes would be involved in the physical decay of $a_1 \to 
3\pi$. The partial widths for $a_1\to\sigma\pi$ found in this model are
similar to those in other extended chiral models\cite{INV88,W90}. In contrast,
the Particle Data Group~\cite{PDG} quotes an experimental upper bound on the
final state $\pi (\pi\pi)_{S}$ of $\sim 0.7 \%$ of the total $a_1$ width of
$\sim 400$ MeV. The strong couplings obtained here suggest that the model may
not be consistent with such a result. However, the situation is far from
clear. The two-stage process $a_1\to\sigma\pi \to 3\pi$ would have to be
integrated over the momentum of the intermediate scalar meson. For momenta
corresponding to a virtual $\sigma$, the $a_1\sigma\pi$ and $\sigma\pi\pi$
couplings may be reduced from their on-shell values. A hint that this may
indeed be the case is provided by the $\sigma\pi\pi$ loop integral, which
vanishes at a total energy of around $800$ MeV. Furthermore, there is also an
amplitude for the $(\pi\pi)_S\pi$ final state originating from a direct,
four-point $a_1 \to 3\pi$ diagram which we have not yet calculated. Although
an estimate of this coupling in the derivative expansion of the  extended NJL
model\cite{Pr94} finds it to be small, it is conceivable that it might
conspire to cancel some of the amplitude due to the intermediate scalar meson.
However, as discussed in Ref.\cite{BOM92}, such a cancellation is not required
by any underlying principle such as chiral symmetry. Finally one should also
note that, as already discussed, leading-order calculations in $1/N_c$ may not
be sufficient to describe a process in which the scalar isoscalar channel
plays a significant role.

The dominant decay mode of the $a_1$ is to $\rho\pi$. Although the parameter
set B does give a credible, broad width, with set A the state seems to be very
narrow. Since the final state has a combined mass of $910$ MeV, the allowed
phase space for the decay is drastically reduced at the model $a_1$ masses as
compared to the empirical mass. Using parameter set A, the $a_1$ mass is only
$946.8$ MeV and so the small decay width of $44$ MeV may simply be a
consequence of the phase-space suppression.

In order to examine whether the $a_1 \rho\pi$ coupling is reasonably well
described by the model, we compare it with the description of the same process
using a phenomenological mesonic Lagrangian. The CCWZ formalism originally
developed in Ref.\cite{CCWZ} is particularly convenient for this purpose since
the $a_1$ mass can be set to any desired value without violating the
constraints of chiral symmetry. A suitable Lagrangian for the comparison is
one obtained by converting the simplest Lagrangian in the massive Yang--Mills
scheme\cite{Mei88} into its CCWZ equivalent and then adjusting the $a_1$
mass.\footnote{A similar conclusion is reached if one starts with the simplest
Lagrangian for $\pi$, $\rho$ and $a_1$ in the hidden-symmetry formalism.}
Details of the relevant change of variables can be found in Ref.\cite{Bir96}:
the appropriate Lagrangian is equation~(5.6) of that paper. We take $f_\pi=93$
MeV, $m_\rho=770$ MeV and $Z^2=1/2$ as the parameters specifying the original
massive Yang--Mills Lagrangian. For the empirical $a_1$ mass of $1230$ MeV,
this gives a broad state of width $490$ MeV. When we use the $a_1$ masses
found in the model with parameter sets A and B, the effective Lagrangian gives
significantly smaller widths, $23$ and $132$ MeV respectively. This suggests
that the small widths found in the model are largely due to the the small
$a_1$ mass rather than any underestimate of the coupling strength.

The amplitude for the decay $a_1\to \rho\pi$ is a mixture of $s$ and $d$-wave
components. In terms of the decay parameters defined in~(\ref{decays}), the
ratio of $d$ to $s$-wave amplitudes is
\begin{equation}
R=-\sqrt{2} \frac{(E_\rho - m_\rho)g_{a_1 \rho\pi} + |\hbox{\bf q}_\rho|^2 
m_{a_1} h_{a_1 \rho\pi}}{(E_\rho + 2m_\rho)g_{a_1 \rho\pi} + |\hbox{\bf 
q}_\rho|^2 m_{a_1} h_{a_1 \rho\pi}}, \label{DSratio}
\end{equation}
where $E_\rho$ and $\hbox{\bf q}_\rho$ are the energy and three-momentum of 
the $\rho$, in the $a_1$ rest frame. This quantity has been determined by the
ARGUS collaboration~\cite{ARGUS} from $\tau$-decay data to be $-0.11 \pm 0.02$.
The effective Lagrangian approach discussed above requires higher-order
couplings in order to obtain a non-zero $h_{a_1 \rho\pi}$ and so this ratio
provides a test of such higher-order effects. From the values for $R$ in
Table~4, we see that, for the parameter set A, the ratio is rather low, but
the value for set B is consistent with the observed one. Overall, set B
provides the better description of both the $a_1$ mass and its hadronic decays.
   
\section{Electromagnetic processes}\label{s6}

Further tests of models where mesons are constructed as $\overline{q}q$
composites are provided by electromagnetic processes and form factors that
probe the internal meson structure. For this it is necessary to specify the 
photon-quark coupling, which requires additional assumptions about the form of 
the nonlocal current, as discussed in Sec.~\ref{s2}. The resultant coupling in 
the present model and the means of calculating the electromagnetic decays of 
the vector mesons were discussed in Subsec.~\ref{curr coupl}. These 
photon-vector-meson couplings are defined by
\begin{eqnarray}
\langle 0 | J^{\mu a} | \rho_s^b \rangle & = & - g_{\rho\gamma} \delta^{ab} 
\epsilon^\mu_s, \nonumber \\
\langle 0 | J^\mu | \omega_s \rangle & = & - g_{\omega\gamma} \epsilon^\mu_s .
\end{eqnarray}                                           
The empirical values, obtained from $\rho\to e^+ e^-$  and  $\omega\to e^+
e^-$\cite{PDG}, are $g_{\rho\gamma}=0.1177$ GeV$^2$ and 
$g_{\omega\gamma}=0.0359$ GeV$^2$. The calculated values for these couplings, 
given in Table~5, are in reasonable agreement with the experimental ones.

Values for the dimensionless quantities $g_V=\frac{m_V^2}{g_{V\gamma}}$ are
also given in Table~5 and may be compared with the results for
$g_{\rho\pi\pi}$ in Table~4. Universal coupling of the $\rho$ would predict
that $g_\rho = g_{\rho\pi\pi}$ (see, for example,\cite{Mei88,Bir96}). This is
clearly violated in the model, although notably less so with parameter set A,
where both of these couplings are closer to the empirical ones.

It is interesting to compare the $\rho\gamma$ coupling with the corresponding
coupling of the $a_1$ to the transverse axial current. A coupling strength
$g_{a_1}$ can be defined analogously to $g_{V\gamma}$,
\begin{equation}
\langle 0 | J_5^{\mu a} | a_{1s}^b  \rangle = - g_{a_1} \delta^{ab} 
\epsilon^\mu_s .
\end{equation}            
Evaluating this as described in Subsec.~\ref{curr coupl}, we obtain values for
$g_{a_1}$ of $0.072$ GeV$^2$ and $0.138$ GeV$^2$ with parameter sets A and B,
respectively. No direct experimental measurement of $g_{a_1}$ exists with
which to make a comparison, but the quantity does appear in Weinberg's sum
rules\cite{W67}. If one assumes complete vector and axial-vector meson
dominance in Weinberg's first and second sum rules, then one gets the
relations
\begin{eqnarray}
\frac{g_{\rho\gamma}^2}{m_\rho^2} - \frac{g_{a_1}^2}{m_{a_1}^2} & = & f_\pi^2 ,
\\
g_{\rho\gamma} & = & g_{a_1} .
\end{eqnarray}
The results of the model for parameter set A are consistent with these
vector-dominance versions of these sum rules, at the $\sim 15\%$ level. In
contrast the results for set B clearly fail to satisfy these relations.

The pion form factor provides a further test of vector-meson dominance (VMD).
This quantity receives contributions from the two kinds of diagram shown in
Fig.~7. The first of these involves a triangular loop, and is often the only
one considered in calculations of the form factor. For a timelike momentum $q$
carried by the external current, the situation is similar to that discussed in
Sec.~\ref{s5} for the triangle diagrams in hadronic decays, with
pseudo-threshold effects appearing at energies beyond twice the real part of
the quark pole.

The other kind of diagram in Fig.~7, which we shall refer to as a
two-body diagram, is produced by the terms $G_1 (i\gamma_5 \otimes i\gamma_5)$ 
and $G_2 (\gamma^\nu \gamma_5 \otimes \gamma_\nu \gamma_5)$ with type-II 
structure~(\ref{II}) in the nonlocal isovector current. The quark propagators 
involved are evaluated at $p \pm \frac{1}{2} q_\pi$, and so pseudo-threshold
effects are not present. The contribution of these diagrams can be written as a
sum over terms, each of which is the product of two loop integrals which are
somewhat similar to those in $J_{PP}, J_{AP}$ or $J_{AA}^L$. Since the path
variable appears in form factors associated with both of the loops, the two
loops are coupled together. The need for similar two-body contributions has
also been noted in the context of models where the four-quark interaction is
dependent only on the relative momentum of the $\overline{q}q$ pair\cite{IBG}.
In a model of that type the analogue of the two-body diagram can be reduced to
a single two-quark loop, where one of the the $\pi qq$ vertex functions is
modified by the photon. Such diagrams make no contribution to the pion charge,
unlike the two-body diagrams in the present model.

The absolute value of the form factor is displayed in Fig.~8, for a parameter
set with $m_0(0)=300$ MeV. Below the $\rho$ meson pole, very little variation 
with $m_0(0)$ is observed in the results when other parameter sets are used. 
Also shown on the figure are experimental data points, obtained from 
Refs.~\cite{A+86,B+73} for the region of spacelike $q^2$ and Ref.\cite{B+85} 
for timelike $q^2$. Below the $\rho$ pole, the model curve is in rather good 
agreement with the data, although its rise is a little shallower. This is 
confirmed by the mean-square pion radius, given in Table~5. The values are 
somewhat smaller than the experimental one \cite{A+86} of $0.439 \pm 0.008$ 
(fm)$^2$.
  
The dashed curve in Fig.~8, labelled VMD, is plotted to test to what
extent the model is consistent with the concept of vector meson dominance. 
Under the assumption of VMD in the photon-pion coupling, the form factor is
\begin{equation}
F_\pi (q^2) = 1 - \frac{g_{\rho\pi\pi}}{g_\rho} \frac{q^2}{q^2 - m_\rho^2}.
\label{VMD ff}
\end{equation}
The ratio of $g_{\rho\pi\pi}$ to $g_\rho$ is underestimated by the model, and 
as a result the VMD form factor rises somewhat more slowly than the data. 
Nonetheless the VMD approximation to the model curve is not a bad one, 
particularly at low $q^2$.

An important check on our calculations, both analytic and numerical, is that  
the pion charge should be unity, $F_\pi(0)=1$. This result has been verified
analytically in the chiral limit\cite{Pla}. It is also satisfied by our 
numerical results, to the accuracy of our integration routines. About a third
of the pion charge comes from the two-body diagram in Fig.~7, demonstrating
the importance of including this contribution.

The individual contributions to the form factor, below the $\rho$ pole, are
shown in Fig.~9. The curve labelled ``bare" is the contribution from the 
triangle diagram with the local $\gamma qq$ coupling (the first diagram
in the representation of the dressed $\gamma qq$ vertex shown in Fig.~5).
The curve labelled ``scalar" comes from the triangle diagram with a nonlocal
coupling (as in the second diagram of Fig.~5) and a loop with a scalar
insertion, given in Eq.~(\ref{gammaqq sc}). This contribution is negligible
over the range of $q^2$ considered. All other contributions are included in
the curve labelled ``vector", corresponding to the part of Eq.~(\ref{gammaqq
full}) that is proportional to $B(q^2)$. Since the transverse $\rho$-meson
propagator is contained in the function $B_2(q^2)$~(\ref{b2}), this
contribution is the dominant one close to the $\rho$ pole. The contribution of
the two-body diagram is relatively small in the vicinity of the $\rho$.
However it varies only slowly with momentum, and so it quite rapidly becomes
important at large spacelike momenta. This is as expected since, for large 
momentum transfer to the pion, the pion vertex functions cut down the
triangle-diagram contributions.

It is interesting to note that, away from the pole, much of the variation with
momentum is controlled by the ``bare" piece of the form factor, which
contributes $\sim 77 \%$ of the mean-square charge radius $\langle r_\pi^2
\rangle$. Although this term has no $\rho$ pole, when added to the ``vector" 
contribution, the result is close to that of the VMD approximation to the 
model. This implies cancellation between the ``bare" contribution and
that of other states above the $\rho$ in the ``vector" piece, to leave the 
$\rho$ pole as the dominant contribution. In contrast, although a similar
mean-square radius has been obtained in an extended NJL model\cite{BM}, most
of that value was accounted for by a diagram with an intermediate $\rho$
meson, the bare photon vertex yielding just 32\%.

Above the $\rho$ pole, the measured form factor is not well described by the
model curve. A possible explanation is suggested by Fig.~10, which breaks the
form factor down into its various contributions for $q^2>m_\rho^2$. There is a
substantial cancellation between the ``bare" and ``vector" contributions and
so the result is likely to be very sensitive to fine details of the model in
this regime. This is borne out by the strong dependence of the results in this
region on the parameter set chosen. It is also responsible for the prominence
of the rather strange structure just above the pseudo-threshold in Fig.~8.
This cancellation between large amplitudes suggests that the results of the
model should not be regarded as reliable in this region.

The use of the conserved current constructed in Sec.~\ref{s2} implicitly 
ensures that the electromagnetic Ward identities are satisfied by the model.
Several examples of these have been already discussed above. Another important
one is the amplitude for the decay $\pi^0 \to\gamma\gamma$, which is an example
of an anomalous process. Such processes require a complete set of quark states
and so present a problem for the usual NJL model, where the use of a regulator 
means that high-energy quark states are discarded. In the model studied here,
the low-energy theorem for $\pi^0 \to\gamma\gamma$ is automatically satisfied
provided that one includes both of the diagrams shown in Fig.~11. As well as
the usual triangle diagram, there is a two-body diagram involving a dressed
$\gamma qq$ vertex~(\ref{gammaqq full}) for one of the photons and a $\gamma
qqqq$ vertex for the other. Details of the calculation are given in the
Appendix.

The finite current quark masses mean that the physical amplitude differs 
slightly from its value in the chiral limit. Numerical results for 
$g_{\pi\gamma\gamma}$, defined from
\begin{equation}
\langle \gamma (q_1) \gamma (q_2) | \pi^0 \rangle = \frac{-2 \alpha_{EM}}{\pi
f_\pi} g_{\pi\gamma\gamma} \epsilon_{\alpha\beta\mu\nu} q_1^\alpha q_2^\beta 
\epsilon_1^{\ast\mu} \epsilon_2^{\ast\nu}
\end{equation}
are given in Table~5. The deviations from $\frac{1}{2}$, the result in
the chiral limit, are small and are consistent with those in the experimental
value, $g_{\pi\gamma\gamma}=0.503 \pm 0.018$\cite{PDG}. 

The related process where one of the photons is off-shell, $\gamma\gamma^*\to
\pi^0$, provides a probe of the structure of the neutral pion. A
corresponding form factor may be defined as
\begin{equation}
F_{\pi\gamma} (q^2) = \frac{\langle \gamma^\ast (q^2) \gamma | \pi^0 
\rangle}{\langle \gamma (q^2=0) \gamma | \pi^0 \rangle}.
\end{equation}                     
As well as older data from the CELLO collaboration\cite{CELLO}, there is now
new data from CLEO\cite{CLEO} on this form factor in the spacelike region,
which extends from $-m_\pi^2$. These are shown in Fig.~12, along with the
results of the model for a parameter set with $m_0(0)=300$ MeV. The model
results are not sensitive to the choice of parameter set and are dominated by
the contribution from the triangle diagram with local photon couplings. They
are consistent with the limited data available and comparable to the results
obtained in other Bethe--Salpeter approaches\cite{KK96,FMRT95}. There are hopes
that further experimental data will become available, since there are plans to
measure the form factor with high precision at Jefferson Lab\cite{JLloi}. Also
displayed on Fig.~12 is the VMD prediction for the form factor, given by
\begin{equation}
F_{\pi\gamma}(q^2)=1 - \frac{2 \pi^2 f_\pi}{g_{\pi\gamma\gamma}} \sum_{V=\rho ,
\omega} \frac{g_{V\pi\gamma}}{g_V} \frac{q^2}{q^2 - m_V^2} ,
\end{equation}
using the model values for the couplings. (The couplings $g_{V \pi\gamma}$,
describing the decays $V \to \pi\gamma$, will be discussed shortly.) In this 
case, the VMD approximation to the model result is a little poorer than it was
for the pion form factor. 

Finally, we consider decays of spin-1 mesons into $\pi\gamma$ final states.
Schematically, the calculation involves triangle and two-body diagrams,
analogous to those of Fig.~11. Just as in the anomalous pion decay the piece
of the nonlocal current which gives rise to the two-body diagrams for $V
\to\pi\gamma$ is the type-I term (\ref{I}) with Dirac structure $G_2
(\gamma^\nu \gamma_5 \otimes \gamma_\nu \gamma_5)$. These two-body diagrams do
not prove to be numerically important in these decays, producing less than $1
\%$ of the total amplitudes. The results for these couplings, as defined by 
\begin{eqnarray}
\langle \gamma (q_1) \pi^0 (q_2) | \omega \rangle & = & i e g_{\omega\pi\gamma}
\epsilon_{\alpha\beta\mu\nu} q_1^\alpha q_2^\beta \epsilon_\gamma^{\ast\mu}
\epsilon_\omega^\nu, \nonumber \\
\langle \gamma (q_1) \pi^b (q_2) | \rho^a \rangle & = & i \delta^{ab} e 
g_{\rho\pi\gamma} \epsilon_{\alpha\beta\mu\nu} q_1^\alpha q_2^\beta 
\epsilon_\gamma^{\ast\mu} \epsilon_\rho^\nu,
\end{eqnarray}
are given in Table~5, along with the corresponding decay widths. In the
isospin symmetric case, there is no $\epsilon^{ab3}$ component to the
$\rho\pi\gamma$ matrix element, so that the decay widths for the charged and
neutral $\rho$ mesons are equal. The model results agree well with the 
experimental values~\cite{PDG}, 
\begin{eqnarray}
\Gamma (\omega \to \pi \gamma) & = & 717 \pm 43 \hbox{keV} , \nonumber \\
\Gamma (\rho^0 \to \pi^0 \gamma) & = & 121 \pm 31 \hbox{keV} , \nonumber \\
\Gamma (\rho^\pm \to \pi^\pm \gamma) & = & 68 \pm 8 \hbox{keV} ,
\end{eqnarray}
the difference between the measured charged and neutral $\rho$-decays not
being considered statistically significant in view of the large error
bars\cite{DAPHNE}.

Extending the $\omega\pi\gamma$ amplitude to off-shell photon momenta, we
compare our model results to the form factor as measured in Ref.\cite{D+81}.
The reaction $\pi^- p \to n\omega \to n\pi^0\mu^+\mu^-$ was used to probe the
form factor in the range from $4m_\mu^2$ to $(m_\omega - m_\pi)^2$. Using a
definition of
\begin{equation}
F_{\omega\pi} (q^2) = \frac{\langle \gamma^\ast (q^2) \pi | \omega 
\rangle}{\langle \gamma (q^2=0) \pi | \omega \rangle} ,
\end{equation}                     
the model results and experimental data are shown in Fig.~13. A parameter set
where $m_0(0)=300$ MeV has again been used in plotting the model results,
which are not sensitive to the set chosen. In common with other theoretical
calculations\cite{BH94} the model is in agreement with the data points at low
$q^2$ but there is a discrepancy with the higher-energy values. It may be that
there is some effect on this form factor due to the tail of the $\rho^\prime$
resonance\cite{L85}. Also, a potentially important missing ingredient in our
calculation is $\omega\phi$ mixing, since a calculation of this form factor
within an SU(3) effective Lagrangian approach\cite{KKW96} has found a
significant dependence on the mixing strength. Improved data would be needed
to draw any firmer conclusions and there are hopes that the experimental
situation will be indeed be clarified by forthcoming experiments at VEPP-2M or
DA$\Phi$NE\cite{DAPHNE}.

In this case, the comparison with the VMD prediction,
\begin{equation}
F_{\omega\pi} (q^2) = 1 - \frac{g_{\omega\rho\pi}}{g_{\omega\pi\gamma}g_\rho}
\frac{q^2}{q^2 - m_\rho^2}, \label{ogpff}
\end{equation}
is not completely straightforward, since the coupling $g_{\omega\rho\pi}$
cannot be calculated on-shell. Nonetheless by extrapolating to the soft-pion
limit (zero pion four-momentum) we can get a reasonable estimate for the
strength at the pole in Eq.~(\ref{ogpff}). The results for $g_{\omega\rho\pi}$
determined in this way are within 20\% of the prediction of universal coupling,
$g_{\omega\rho\pi}=g_{\omega\pi\gamma}g_\rho$. The results for the form
factor plotted in Fig.~13 again show good agreement between the full 
calculation and the VMD approximation.

In the amplitude for the decay $a_1^\pm \to \pi^\pm \gamma$, the two-body
diagram involves the terms in the nonlocal current $G_1 (i\gamma_5 \otimes
i\gamma_5)$ and $G_2 (\gamma^\nu \gamma_5 \otimes \gamma_\nu \gamma_5)$ which
have a type-II structure (\ref{II}). Since the $a_1$ mass lies above the
pseudo-threshold for parameter set B, in that case both the triangle and
two-body diagrams require residue contributions. Gauge invariance imposes the
following structure on the amplitude,
\begin{equation}
\langle \gamma (q_1) \pi^b (q_2) | a_1^a \rangle = i \epsilon^{ab3} e g_{a_1 
\pi\gamma} \left[ \epsilon_{a_1}\cdot \epsilon_\gamma^\ast + \frac{2 
(q_2\cdot\epsilon_{a_1}) (q_2\cdot\epsilon^\ast_\gamma)}{(m_{a_1}^2 - m_\pi^2)} 
\right] .
\end{equation}
The values calculated for $g_{a_1 \pi\gamma}$, and the corresponding decay
widths, are presented in Table~5. With parameter set A the dressed scalar
piece of the nonlocal photon coupling and the two-body diagrams make
relatively small contributions; for set B, these pieces are rather larger in
magnitude but largely cancel with each other. The final results for all choices
of parameters are much smaller than the experimental measurement\cite{Z+84} of
$640 \pm 246$ keV.
 	
\section{Conclusions}\label{s7}
We have investigated extended versions of the model proposed in
Ref.\cite{BB95} with nonlocal, four-fermion interactions, based on the
separable form (\ref{separable}) suggested by instanton-liquid
studies\cite{DP}. The interaction form factors ensure the convergence of all
quark-loop integrals and, for most of the acceptable range of parameters, also
lead to quark confinement~\cite{BB95}. The familiar problems of the NJL model
are therefore eliminated whilst retaining much of the simplicity of that
approach.

Quark confinement is particularly important in applying such a model to heavier
mesons than the pion, since it avoids a threshold for $\overline{q}q$
production occurring at an inconveniently low energy, as in the NJL model. This
feature of the nonlocal approach makes it especially well suited as the basis
for the extended model that we study here, with interactions that bind vector
and axial-vector mesons.

The analytic structure of the quark propagator, with poles at complex 
momenta, means that a scheme has to be specified for continuing amplitudes to
Minkowski space. We have used the scheme suggested by Lee and Wick\cite{LW69}
and Cutkosky {\it et al.}\cite{CLOP}, which leads to nonanalytic behaviour of
the meson propagators above a pseudo-threshold. This behaviour ensures that no
poles appear in the longitudinal vector channels. It also provides restrictions
on both the admissible range of model parameters and the region of
applicability of the model.

In order to calculate the pion decay constant, which sets the basic scale for
the model, as well as electromagnetic couplings, we need to construct conserved 
currents that are consistent with our nonlocal interaction. This has been done
using the Noether-like method of Ref.\cite{BB95}. Ward identities related to
the current conservation follow automatically and several have been
analytically checked, including the Gell-Mann--Oakes--Renner relation and
the low-energy theorem for $\pi^0\to\gamma\gamma$ in the chiral limit. The 
latter involves the axial anomaly, which has formed a long-standing problem
for the usual NJL model with its cut-offs on the quark propagators.

Working at tree level in terms of mesons (leading order in $1/N_c$) and to all 
orders in momentum, we have calculated the meson masses and various strong 
decay rates. When the $\rho$ mass is used to fix the strength of the relevant
four-quark vertex, we find that the $\rho$-meson width is reasonably well
described. In particular the model provides an improvement on the
underestimated value obtained in the extended NJL model\cite{rhodecay}.
As in the NJL model, a light $\sigma$ is found, with a mass similar to that
found in a number of analyses of $\pi\pi$ scattering\cite{sigma}. The 
calculated mass of the $a_1$ meson is somewhat lighter than the observed
value. By cutting down the available phase-space, this also means that the 
model gives too small a width for the decay $a_1\to\rho\pi$.

Various electromagnetic quantities have also been calculated in this model.
The decay rates found in the model are in reasonably good agreement with the
observed ones, except for the case of $a_1\to\pi\gamma$. The electromagnetic
form factor of the pion agrees well with the data, at least below the $\rho$
pole. The form factors for $\gamma\gamma^*\to \pi$ and $\omega\to\pi\gamma^*$
are also in agreement with the, admittedly rather limited, data currently
available. We have also compared our results for electromagnetic form factors
with vector-dominance formulae using on-shell couplings as calculated in the
model. Although diagrams where the photon couples directly to the quarks make a
significant contribution, they combine with diagrams that involve an
intermediate vector meson to give results that are very close to those of VMD,
at least at low momenta. The model is thus able to provide some support for the
idea of vector dominance in photon-meson interactions.

Overall the level of agreement with observed meson properties is satisfactory,
given the simple nature of the model and the fact that important effects have
not been included at the current level of approximation. It is desirable to
extend the current work to include meson-loop effects ($1/N_c$ corrections)
since these could be significant. In particular, they may be important for the
scalar isoscalar channel, where the $\sigma$ meson is very strongly coupled to
two-pion states and where there are some indications that our leading-order
calculations do not agree with phenomenological analyses.

\section*{Acknowledgements}
MCB acknowledges helpful discussions with W. Broniowski. This work is
supported by the EPSRC and PPARC.

\appendix
\section{$\pi^{0} \to \gamma\gamma$}

We outline here the various contributions to the $\pi^0 \to \gamma\gamma$
amplitude in this model. We work in the chiral limit, where the axial anomaly
requires that $g_{\pi\gamma\gamma} = \frac{1}{2}$. The relevant diagrams are
shown in Fig.~11. Consider first the simpler case where vector meson degrees
of freedom are not present in the model, setting $G_2=G_3=0$. The triangle
diagrams where both of the photons couple to the local current, $\gamma_\mu$,
contribute
\begin{eqnarray}
4i \epsilon_{\alpha\beta\mu\nu} q_1^\alpha q_2^\beta \epsilon_1^{\ast\mu} 
\epsilon_2^{\ast\nu} e^2 g_{\pi qq} \int \frac{d^4 p}{(2\pi)^4} 
\frac{f^2(p)}{[p^2 - m_0^2(p^2)]^3} \left( -2m_0(p^2) + 4 m_0^\prime(p^2)
\frac{(p\cdot q)^2}{q^2}\right. \nonumber\\
\left.+ 4 m_0^\prime(p^2) \frac{[p\cdot(q_1 - q_2)]^2}
{(q_1 - q_2)^2} \right) , \label{VV anom}
\end{eqnarray}
where the prime denotes a derivative with respect to $p^2$. The 
contribution of diagrams where one of the photons couples to the nonlocal
piece of the current, with a scalar loop insertion (\ref{gammaqq sc}), is
\begin{equation}
4i \epsilon_{\alpha\beta\mu\nu} q_1^\alpha q_2^\beta \epsilon_1^{\ast\mu} 
\epsilon_2^{\ast\nu} e^2 g_{\pi qq} \int \frac{d^4 p}{(2\pi)^4} \frac{f^2(p) 
4 m_0^\prime (p^2)}{[p^2 - m_0^2(p^2)]^3} \left( p^2 - \frac{(p\cdot q)^2}{q^2}
- \frac{[p\cdot(q_1 - q_2)]^2}{(q_1 - q_2)^2} \right) . \label{VS anom}        
\end{equation}
Diagrams with a nonlocal coupling at both photon vertices do not contribute to
the amplitude because the resulting Dirac trace vanishes. Converting the
integral in the sum of (\ref{VV anom}) and (\ref{VS anom}) into Euclidean
space, and changing variable to
\begin{equation}
t= \frac{m_0^2 (p_E^2)}{p_E^2} ,
\end{equation}
leads to 	
\begin{equation}
- 2 \epsilon_{\alpha\beta\mu\nu} q_1^\alpha q_2^\beta \epsilon_1^{\ast\mu}
\epsilon_2^{\ast\nu} \frac{\alpha_{EM}}{\pi} \frac{g_{\pi qq}}{m_0(0)} 
\int_0^\infty \frac{dt}{(1 + t)^3} .
\end{equation}
The low-energy theorem now follows by invoking the analogue of the
Goldberger--Treiman relation\cite{GT} in this model,
\begin{equation}
g_{\pi qq} = \frac{m_0 (0)}{f_\pi} , \label{gt}
\end{equation}
which was derived in Ref.\cite{BB95} by considering $f_\pi$ in the chiral
limit.\footnote{In the notation of Ref.\cite{BB95} $g_{\pi qq}$ is 
$Z_\pi^{\frac{1}{2}}$.}

In the extended model with vector mesons, $g_{\pi qq}$ is affected by the 
pseudoscalar-axial mixing induced by the $G_2$ coupling (\ref{pi
couplings},\ref{piondet}). However, the form of $f_\pi$ is also modified, as
discussed in Subsec.~\ref{curr coupl}, and hence Eq.(\ref{gt}) remains valid
in the chiral limit\cite{Pla}. In addition the photons are on-shell and so we
need only consider the bare $\gamma qq$ vertex and the nonlocal
piece~(\ref{gammaqq sc}) with a scalar one-quark loop, as also discussed in
Subsec.~\ref{curr coupl}. The sum of the contributions~(\ref{VV anom})
and~(\ref{VS anom}) therefore yields $g_{\pi\gamma\gamma}= \frac{1}{2}$, just
as in the model without vector mesons. 

This feature is in agreement with the analysis of Ref.~\cite{BHK94}, where it
was shown that for a quark propagator without wave function renormalization,
then the anomaly is saturated by taking only the leading part of the pion
Bethe--Salpeter amplitude and the dressed $\gamma qq$ structures subject to
the Ward identity for that vertex. The statement is non-trivial because terms
in the pion amplitude that are linear in momentum can contribute to the decay
amplitude, even in the chiral limit. For instance, the $\qs\gamma_5$ term 
which appears in $V_\pi(q)$ for this model gives rise to additional
triangle-diagram contributions. From such diagrams with two local photon
vertices, one finds
\begin{eqnarray}
-4i \epsilon_{\alpha\beta\mu\nu} q_1^\alpha q_2^\beta \epsilon_1^{\ast\mu}
\epsilon_2^{\ast\nu} e^2 \frac{\widetilde{g}_{\pi qq}}{m_\pi} \int \frac{d^4 
p}{(2\pi)^4} \frac{f^2(p)}{[p^2 - m_0^2(p^2)]^3} \left( -2p^2 - 2m_0^2(p^2) 
+ 4 \frac{(p\cdot q)^2}{q^2} \right.\nonumber\\
\left. + 4 \frac{[p\cdot(q_1 - q_2)]^2}{(q_1 - q_2)^2} 
\right) . \label{VV anom2}
\end{eqnarray}
Since $\widetilde{g}_{\pi qq}$ is of $\O (m_\pi)$ in the chiral expansion,
this expression is of $\O (1)$. Similar diagrams with one local and one
nonlocal vertex give
\begin{equation}
-4i \epsilon_{\alpha\beta\mu\nu} q_1^\alpha q_2^\beta \epsilon_1^{\ast\mu} 
\epsilon_2^{\ast\nu} e^2 \frac{\widetilde{g}_{\pi qq}}{m_\pi} \int \frac{d^4 
p}{(2\pi)^4} \frac{f^2(p) 8 m_0(p^2) m_0^\prime(p^2)}{[p^2 - m_0^2(p^2)]^3} 
\left( p^2 - \frac{(p\cdot q)^2}{q^2} - \frac{[p\cdot(q_1 - q_2)]^2}
{(q_1 - q_2)^2} \right) . \label{VS anom2}
\end{equation}
(The diagrams with two nonlocal vertices again have a vanishing Dirac trace.)

Since the sum of (\ref{VV anom2}) and (\ref{VS anom2}) is non-zero, there must
be some contribution that cancels them in the full amplitude for the anomalous
process. The relevant piece arises from the two-body diagram, which is 
displayed in Fig.~11. Terms in the nonlocal vector current of the form
$G_2(\gamma^\nu \gamma_5 \otimes \gamma_\nu \gamma_5)$ with a type-I~(\ref{I}) 
structure are responsible for the $\gamma qqqq$ vertex. (Note that for 
simplicity we have suppressed isospin factors.) This diagram factorizes into 
two separate loop integrals, the loop between the two photons giving rise
to the $\epsilon_{\alpha\beta\mu\nu}$ factor. The other loop is simply a
linear combination of the familiar integrals $J_{AP}(m_\pi^2)$ and
$J_{AA}^L(m_\pi^2)$ from Subsec.~\ref{meson props}. This combination can be
simplified by recalling the definitions of the pion-quark coupling constants
in (\ref{pi couplings}). In the chiral limit, the contribution of this diagram 
can be written as
\begin{equation}
-4i \epsilon_{\alpha\beta\mu\nu} q_1^\alpha q_2^\beta \epsilon_1^{\ast\mu}
\epsilon_2^{\ast\nu} e^2 \frac{\widetilde{g}_{\pi qq}}{m_\pi} \int \frac{d^4 
p}{(2\pi)^4} \frac{4 f(p)f^\prime(p)}{[p^2 - m_0^2(p^2)]^2} \left( p^2 - 
\frac{(p\cdot q)^2}{q^2} - \frac{[p\cdot(q_1 - q_2)]^2}{(q_1 - q_2)^2} 
\right) .
\end{equation}
Converting to Euclidean space and integrating by parts, this expression can
be shown to cancel exactly with the sum of (\ref{VV anom2}) and (\ref{VS
anom2}), demonstrating that the low-energy theorem for $\pi^0\to\gamma\gamma$
holds in the extended model.

\newpage
\section*{Tables}

\begin{center}
\begin{tabular}{| c | c | c || c | c | c |}
\hline
Parameter\ &\ Set A\ &\ Set B\ &\ Parameter\ &\ Set A\ &\ Set B\\
\hline
$m_0(0)$(MeV)&$280$&$320$&$m(0)$(MeV)&$326$&$370$\\
$m_c$(MeV)&$8.4$&$11.0$&$\Lambda$(MeV)&$995$&$846$\\
$G_1$(GeV$^{-2}$)&$37.1$&$57.6$&$G_2$(GeV$^{-2}$)&$-5.70$&$-6.53$\\
$G_3$(GeV$^{-2}$)&$-5.20$&$-5.86$&$G_4$(GeV$^{-2}$)&$-0.80$&$-4.14$\\
$G_5$(GeV$^{-2}$)&$2.57$&$4.76$& & &\\
\hline                         
\end{tabular}
\end{center}
\noindent Table 1. Values of the model parameters for sets A and B, fitted as
discussed in the text. Also shown is the dynamical quark mass $m(0)$. 
\vspace{2 cm}

\begin{center}
\begin{tabular}{|c | c|}
\hline
Set A\ &\ Set B\\
\hline
$\pm 496 \pm 130i$\ &\ $\pm 404 \pm 257i$\\
$\pm 1168 \pm 790i$\ &\ $\pm 962 \pm 702i$\\
$\pm 1488 \pm 1155i$\ &\ $\pm 1242 \pm 1005i$\\
$\pm 1742 \pm 1436i$\ &\ $\pm 1463 \pm 1240i$\\
\hline
\end{tabular}
\end{center}
\noindent Table 2. Positions of the lowest four sets of poles in the quark 
propagator.
\vspace{2 cm}

\begin{center}
\begin{tabular}{| c | c | c | c | c | c | c |}
\hline
 &\multicolumn{3}{|c|}{Set A} & \multicolumn{3}{|c|}{Set B} \\  
\hline
Particle\ &\ Mass\ &\ $g_{iqq}$\ &\ $\widetilde{g}_{iqq}$\ &\ Mass\ &
\ $g_{iqq}$\ &\ $\widetilde{g}_{iqq}$\\
\hline
$\pi$\ &\ Fit\ &\ $3.44$\ &\ $0.0739$\ &\ Fit\ &\ $3.91$\ &\ $0.0715$\\
$\sigma$\ &\ $443.2$\ &\ $3.51$\ &\ --\ &\ $465.8$\ &\ $4.06$\ &\ --\\
$\rho$\ &\ Fit\ &\ $1.12$\ &\ --\ &\ Fit\ &\ $1.11$\ &\ --\\
$a_1$\ &\ $946.8$\ &\ $1.13$\ &\ --\ &\ $1061.5$\ &\ $2.27$\ &\ --\\
$\omega$\ &\ Fit\ &\ $1.07$\ &\ --\ &\ Fit\ &\ $1.05$\ &\ --\\
$f_1$\ &\ Fit\ &\ $0.89$\ &\ --\ &\ Fit\ &\ $2.51$\ &\ --\\
$a_0$\ &\ Fit\ &\ $0.75$\ &\ --\ &\ Fit\ &\ $1.71$\ &\ --\\
$\eta^\star$\ &\ $874.9$\ &\ $0.83$\ &\ $0.190$\ &\ $899.4$\ &\ $2.36$\ &\ 
$1.448$\\
\hline
\end{tabular}
\end{center}
\noindent Table 3. Meson masses (in MeV) and couplings of the mesons to
quarks. The couplings $g_{iqq}$ and $\widetilde{g}_{iqq}$ are defined by the
equations~(\ref{meson couplings}) and~(\ref{pi couplings}).
\vspace{2 cm}

\begin{center}
\begin{tabular}{| c | c | c | c | c |}
\hline
 & \multicolumn{2}{|c|}{Set A} & \multicolumn{2}{|c|}{Set B}\\  
\hline
Coupling\ &\ Value\ &\ Width(MeV)\ &\ Value\ &\ Width(MeV)\\
\hline
$g_{\sigma\pi\pi}$(MeV)\ &\ $1438$\ &\ $108.0$\ &\ $1625$\ &\ $135.1$\\ 
$g_{\rho\pi\pi}$\ &\ $5.52$\ &\ $126.0$\ &\ $5.26$\ &\ $114.0$\\ 
$g_{a_1 \sigma\pi}$\ &\ $10.65$\ &\ $74.0$\ &\ $11.77$\ &\ $116.4$\\ 
$g_{a_1 \rho\pi}$(MeV)\ &\ $2174$\ &\ $44.0$\ &\ $4604$\ &\ $376.2$\\ 
$h_{a_1 \rho\pi}$(GeV$^{-1}$)\ &\ $18.19$\ &\ --\ &\ $10.87$\ &\ --\\ 
$R$\ &\ $-0.048$\ &\ --\ &\ $-0.087$\ &\ --\\
\hline
\end{tabular}
\end{center}
\noindent Table 4. The meson couplings, as defined in equation~(\ref{decays}),
together with the corresponding partial widths. $R$ is the ratio of the $d$- to
$s$-wave amplitudes defined in (\ref{DSratio}).
\vspace{2cm}

\begin{center}
\begin{tabular}{| c | c | c || c | c | c |}
\hline
&\ Set A\ &\ Set B\ &\ &\ Set A\ &\ Set B\\    
\hline
Quantity\ &\ Value\ &\ Value\ &\ Quantity\ &\ Value\ &\ Value\\
\hline
$g_{\rho\gamma}$(GeV$^2$)\ &\ $0.0889$\ &\ $0.0773$\ &\ $g_\rho$\ &\ $6.67$\ &\ 
$7.67$\\
$g_{\omega\gamma}$(GeV$^2$)\ &\ $0.0308$\ &\ $0.0265$ &\ $g_\omega$\ &\ 
$19.92$\ &\ $23.12$\\
$\langle r_\pi^2 \rangle$(fm$^2$)\ &\ $0.346$\ &\ $0.344$\ &\ -- &\ -- &\ --\\
$g_{\pi\gamma\gamma}$\ &\ $0.505$\ &\ $0.501$\ &\ -- &\ -- &\ --\\ 
$g_{\omega\pi\gamma}$(GeV$^{-1}$)\ &\ $-2.29$ &\ $-2.25$ &\ $\Gamma(\omega \to
\pi\gamma)$(keV)\ &\ $692$\ &\ $669$\\ 
$g_{\rho\pi\gamma}$(GeV$^{-1}$)\ &\ $-0.755$ &\ $-0.707$ &\ $\Gamma(\rho \to
\pi\gamma)$(keV)\ &\ $71.6$\ &\ $62.7$\\ 
$g_{a_1 \pi\gamma}$(MeV)\ &\ $140.2$ &\ $201.5$ &\ $\Gamma(a_1 \to \pi\gamma)$
(keV)\ &\ $24.7$\ &\ $45.7$\\ 
\hline
\end{tabular}
\end{center}
\noindent Table 5. Electromagnetic couplings of mesons.

\newpage
\section*{Figures}

\begin{picture}(370,100)
\ArrowLine(35,30)(75,30)
\BCirc(85,30){10}
\ArrowLine(95,30)(135,30)
\Text(85,30)[c]{$S_F$}
\Text(150,30)[c]{=}
\ArrowLine(165,30)(205,30)
\Text(220,30)[c]{+}
\ArrowLine(235,30)(275,30)
\BCirc(285,30){10}
\Text(285,30)[c]{$G_1$}
\ArrowLine(295,30)(335,30)
\BCirc(285,70){10}
\Text(285,70)[c]{$S_F$}
\ArrowArc(285,50)(-22,-65,65)
\ArrowArc(285,50)(-22,115,-115)
\end{picture}

\noindent Fig.~1. Schwinger--Dyson equation for the quark propagator in the 
ladder approximation.
\vspace{2 cm}

\begin{picture}(370,105)
\ArrowLine(50,52)(20,82)
\ArrowLine(20,22)(50,52)
\ArrowLine(80,82)(50,52)
\ArrowLine(50,52)(80,22)
\BCirc(50,52){10}
\Text(50,52)[c]{$T$}
\Text(97.5,52)[c]{=}
\ArrowLine(115,22)(145,52)
\ArrowLine(145,52)(115,82)
\ArrowLine(145,52)(175,22)
\ArrowLine(175,82)(145,52)
\BCirc(145,52){10}
\Text(145,52)[c]{$G$}
\Text(192.5,52)[c]{+}
\ArrowLine(210,22)(240,52)
\ArrowLine(240,52)(210,82)
\ArrowArc(280,76)(46,-150,-90)
\ArrowArc(280,76)(46,-90,-30)
\ArrowArc(280,28)(46,30,90)
\ArrowArc(280,28)(46,90,150)
\ArrowLine(320,52)(350,22)
\ArrowLine(350,82)(320,52)
\BCirc(240,52){10}
\Text(240,52)[c]{$G$}
\BCirc(320,52){10}
\Text(320,52)[c]{$T$}
\BCirc(280,74){10}
\Text(280,74)[c]{$S_F$}
\BCirc(280,30){10}
\Text(280,30)[c]{$S_F$}
\end{picture}

\noindent Fig.~2. Bethe--Salpeter equation for $\overline{q}q$ scattering in
the ladder approximation.
\vspace{2 cm}

\begin{picture}(370,100)
\LongArrow(185,0)(185,100)
\LongArrow(105,50)(265,50)
\BBoxc(155,90)(4,4)
\BBoxc(155,60)(4,4)
\Vertex(155,10)2
\Vertex(155,40)2
\BBoxc(215,90)(4,4)
\BBoxc(215,60)(4,4)
\Vertex(215,10)2
\Vertex(215,40)2
\ArrowLine(105,50)(149,50)
\Line(149,50)(149,60)
\CArc(155,60)(6,0,180)
\Line(161,60)(161,58)
\ArrowLine(161,58)(149,42)
\Line(149,42)(149,40)
\CArc(155,40)(6,180,270)
\ArrowLine(155,34)(215,34)
\CArc(215,40)(6,270,0)
\Line(221,40)(221,42)
\ArrowLine(221,42)(209,58)
\Line(209,58)(209,60)
\CArc(215,60)(6,0,180)
\Line(221,60)(221,50)
\ArrowLine(221,50)(265,50)
\Text(260,95)[c]{$p_4$}
\LongArrow(145,85)(145,95)
\LongArrow(145,55)(145,65)
\LongArrow(145,45)(145,35)
\LongArrow(145,15)(145,5)
\LongArrow(225,85)(225,95)
\LongArrow(225,55)(225,65)
\LongArrow(225,45)(225,35)
\LongArrow(225,15)(225,5)
\end{picture}

\noindent Fig.~3. Deformed integration contour in $p_4$ plane, beyond the pinch
point. The open boxes denote the poles of the $p_-$ propagator and the filled
circles those of the $p_+$ propagator. The arrows indicate the directions in
which these poles move as $q_0$ increases.

\begin{picture}(370,90)
\Line(40,29)(80,29)
\Line(40,31)(80,31)
\ArrowArc(100,42)(-23,30,150)
\ArrowArc(100,18)(-23,-150,-30)
\BCirc(80,30){10}
\Text(80,30)[c]{$V$}
\Photon(120,30)(160,30){3}{3}
\Line(210,29)(250,29)
\Line(210,31)(250,31)
\ArrowArc(270,42)(-23,30,150)
\ArrowArc(270,18)(-23,-150,-30)
\BCirc(250,30){10}
\Text(250,30)[c]{$V$}
\Photon(290,30)(330,30){3}{3}
\ArrowArc(307,47)(-24,-45,45)
\ArrowArc(273,47)(-24,135,-135)
\Vertex(290,30){3}
\end{picture}

\noindent Fig.~4. Coupling of particle to an external current. $V$ denotes the 
vertex function (\ref{vertex fn}).

\begin{picture}(370,280)
\ArrowLine(60,220)(100,180)
\ArrowLine(20,180)(60,220)
\Photon(60,220)(60,260){3}{3}
\BCirc(60,220){10}
\Text(60,220)[c]{$\Gamma$}
\Text(122.5,220)[c]{=}
\ArrowLine(185,220)(225,180)
\ArrowLine(145,180)(185,220)
\Photon(185,220)(185,260){3}{3}
\Text(247.5,220)[c]{+}
\ArrowLine(310,220)(350,180)
\ArrowLine(270,180)(310,220)
\Photon(310,220)(310,260){3}{3}
\ArrowArc(327,237)(-24,45,135)
\ArrowArc(327,203)(-24,-135,-45)
\Vertex(310,220){3}
\Text(35,60)[c]{+}
\Photon(110,100)(110,140){3}{3}
\ArrowArc(122,80)(-23,-60,60)
\ArrowArc(98,80)(-23,120,-120)
\ArrowLine(110,60)(150,20)
\ArrowLine(70,20)(110,60)
\BCirc(110,60){10}
\Text(110,60)[c]{$T$}
\Text(185,60)[c]{+}
\Photon(260,100)(260,140){3}{3}
\ArrowArc(272,80)(-23,-60,60)
\ArrowArc(248,80)(-23,120,-120)
\ArrowArc(277,117)(-24,45,135)
\ArrowArc(277,83)(-24,-135,-45)
\ArrowLine(260,60)(300,20)
\ArrowLine(220,20)(260,60)
\BCirc(260,60){10}
\Text(260,60)[c]{$T$}
\Vertex(260,100){3}
\end{picture}

\noindent Fig.~5. The dressed $\gamma qq$ vertex. $T$ denotes the
$\overline{q}q$ scattering matrix in either the transverse or longitudinal 
vector channel.

\begin{figure}
\center{\psfig{file=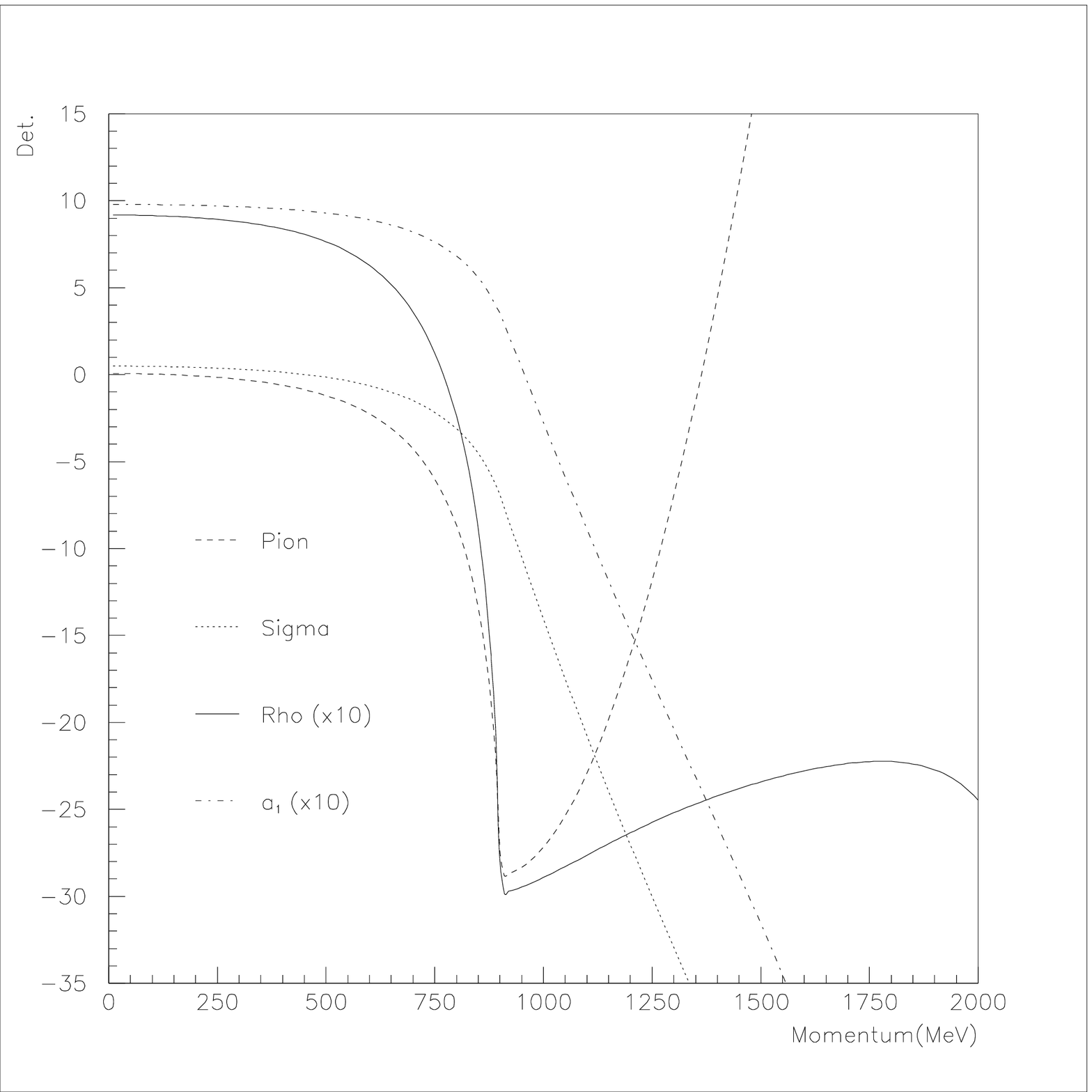,height=6in}}
\end{figure}
	
\noindent Fig.~6. Denominator of the propagator in the sigma channel, $1 - G_1 
J_{SS}$, and the pion determinant, (\ref{piondet}), as functions of the 
timelike meson momentum. The denominators of the $\rho$ and $a_1$ propagators, 
$1 - G_2 J_{VV,AA}^T(q)$, are also displayed, scaled by a factor of $10$.

\newpage
	
\begin{picture}(370,130)
\Line(10,29)(50,29)
\Line(10,31)(50,31)
\ArrowLine(130,30)(50,30)
\Line(130,29)(170,29)
\Line(130,31)(170,31)
\ArrowLine(90,70)(130,30)
\ArrowLine(50,30)(90,70)
\Photon(90,70)(90,110){3}{3}
\BCirc(50,30){10}
\Text(50,30)[c]{V$_\pi$}
\BCirc(130,30){10}
\Text(130,30)[c]{$\overline{\hbox{V}}_\pi$}
\BCirc(90,70){10}
\Text(90,70)[c]{$\Gamma$}
\Line(200,29)(240,29)
\Line(200,31)(240,31)
\ArrowArc(260,42)(-23,30,150)
\ArrowArc(260,18)(-23,-150,-30)
\ArrowArc(300,42)(-23,30,150)
\ArrowArc(300,18)(-23,-150,-30)
\Line(320,29)(360,29)
\Line(320,31)(360,31)
\Photon(280,30)(280,110){3}{6}
\Vertex(280,30){3}
\BCirc(240,30){10}
\Text(240,30)[c]{V$_\pi$}
\BCirc(320,30){10}
\Text(320,30)[c]{$\overline{\hbox{V}}_\pi$}
\end{picture}

\noindent Fig.~7. Spacelike pion form factor. There is another similar 
triangle diagram where the photon couples to the anti--quark. 

\newpage

\begin{figure}
\center{\psfig{file=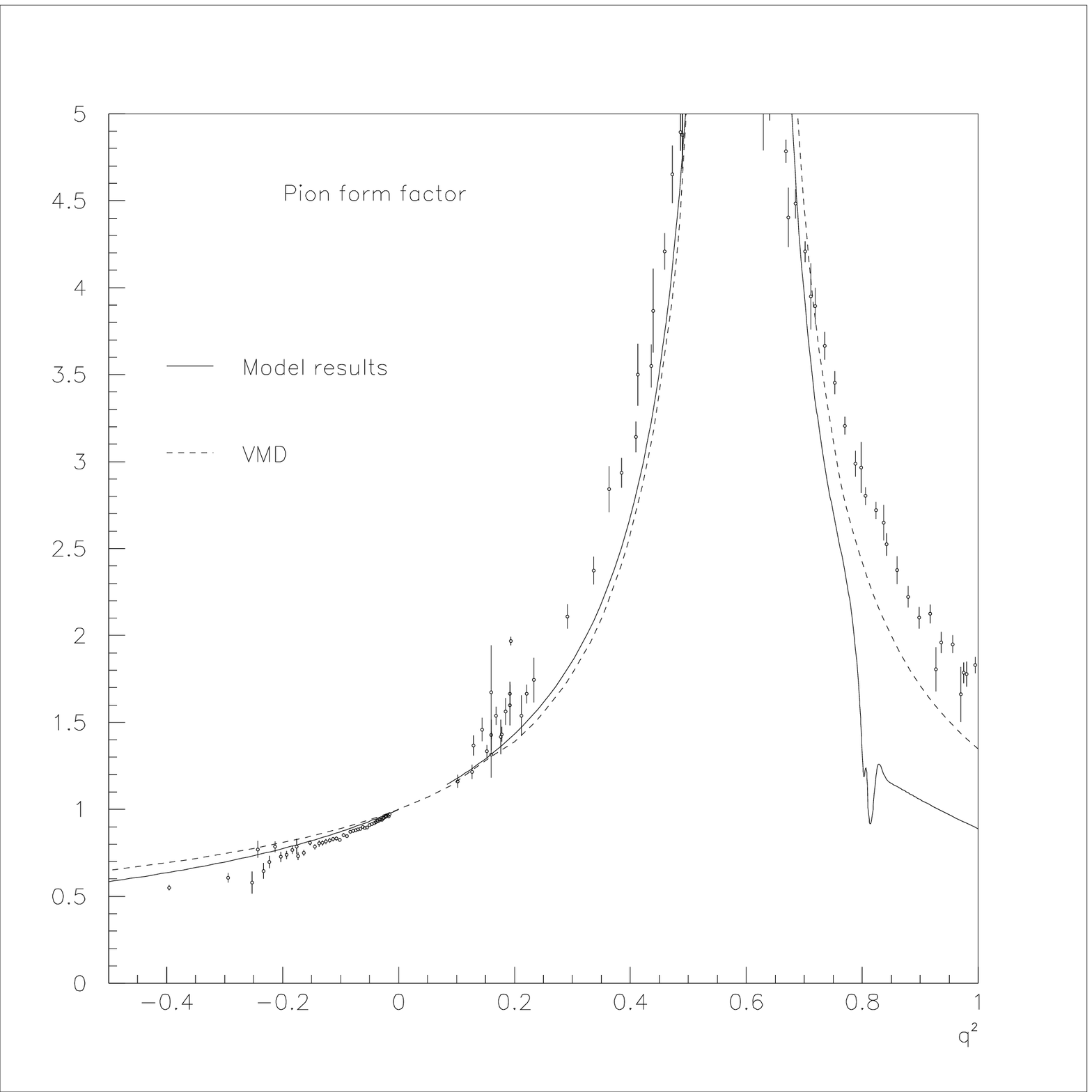,height=6in}}
\end{figure}
	
\noindent Fig.~8. Absolute value of the pion form factor, plotted against 
$q^2$ in GeV$^2$. The solid line is the model result; the dashed line is the
VMD approximation to it. At the level of approximation used the form factor
has a pole at the $\rho$-meson mass. The data points are from
Refs.\cite{A+86,B+73,B+85}.

\begin{figure}
\center{\psfig{file=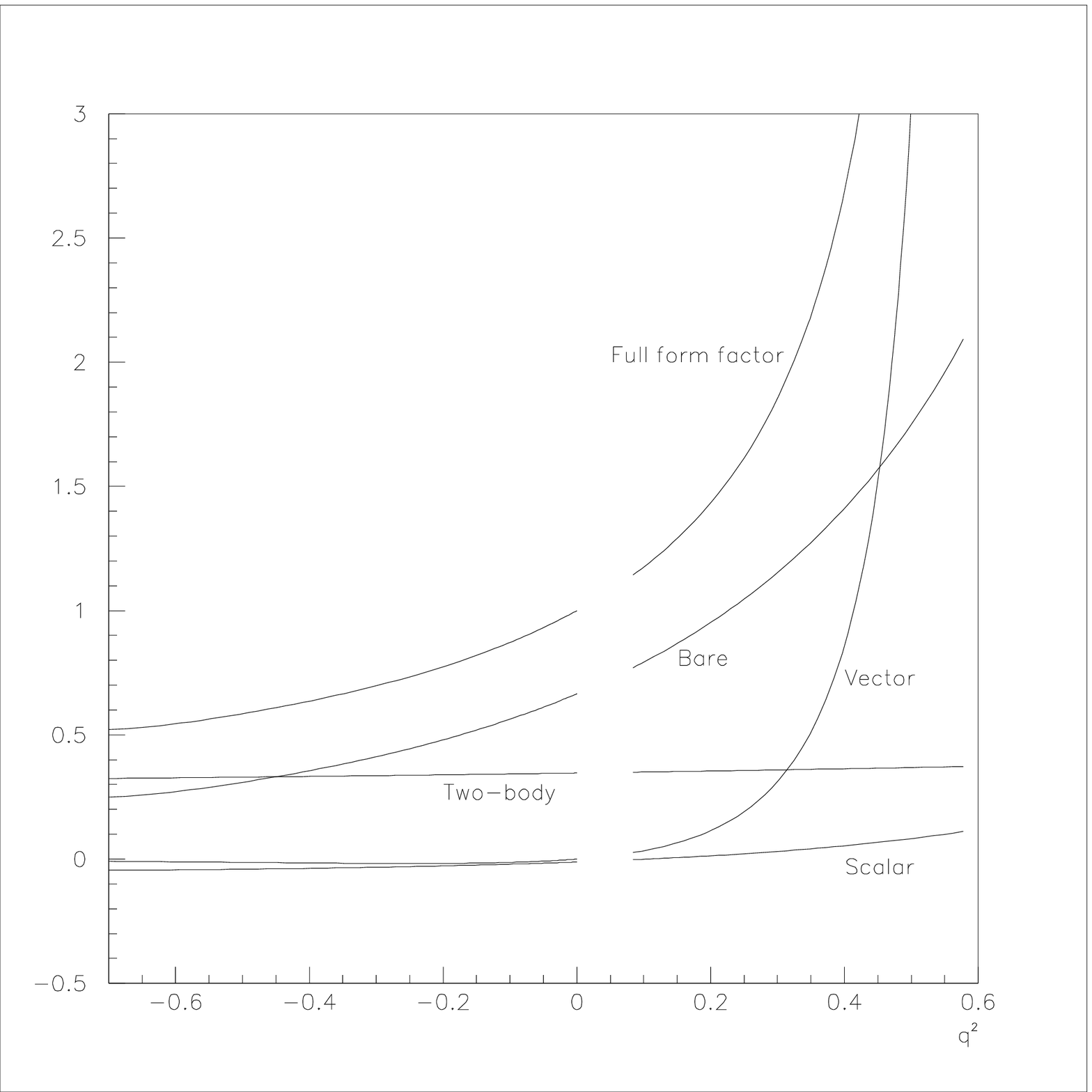,height=6in}}
\end{figure}

\noindent Fig.~9. Various contributions to the pion form factor, below the 
$\rho$ pole, plotted against $q^2$ in GeV$^2$. The different contributions are 
defined in the text. 

\begin{figure}
\center{\psfig{file=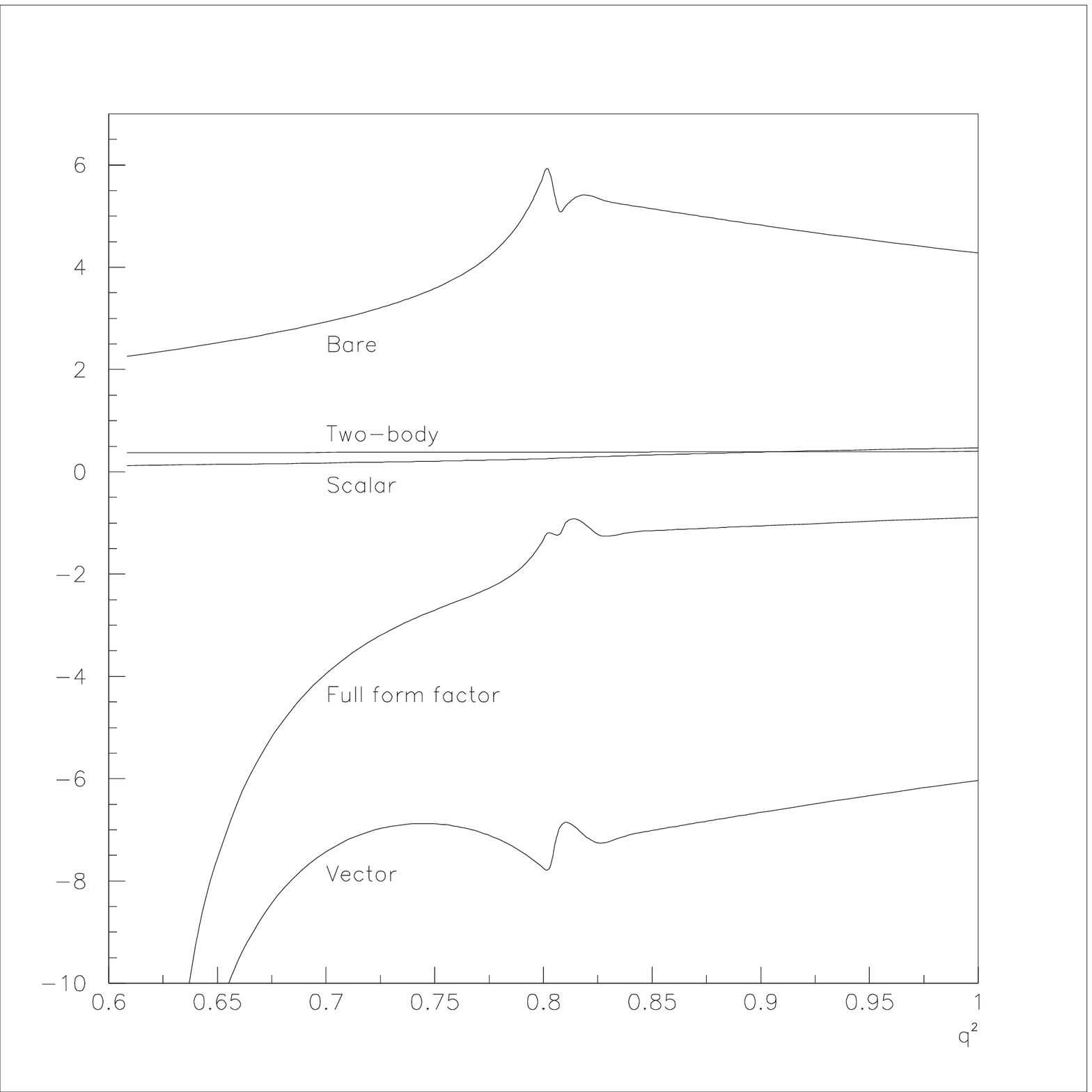,height=6in}}
\end{figure}

\noindent Fig.~10. Various contributions to the timelike pion form factor, 
above the $\rho$ pole, plotted against $q^2$ in GeV$^2$. The different 
contributions are defined in the text. 

\newpage

\begin{picture}(370,120)
\Line(40,59)(80,59)
\Line(40,61)(80,61)
\ArrowLine(80,60)(120,100)
\Photon(120,100)(160,100){3}{3}
\ArrowLine(120,100)(120,20)
\Photon(120,20)(160,20){3}{3}
\ArrowLine(120,20)(80,60)
\BCirc(80,60){10}
\Text(80,60)[c]{V$_\pi$}
\BCirc(120,100){10}
\Text(120,100)[c]{$\Gamma_a$}
\BCirc(120,20){10}
\Text(120,20)[c]{$\Gamma_b$}
\Line(210,59)(250,59)
\Line(210,61)(250,61)
\ArrowArc(270,72)(-23,30,150)
\ArrowArc(270,48)(-23,-150,-30)
\Photon(290,60)(330,60){3}{3}
\Vertex(290,60){3}
\ArrowArc(302,40)(-23,-60,60)
\ArrowArc(278,40)(-23,120,-120)
\Photon(290,20)(330,20){3}{3}
\BCirc(250,60){10}
\Text(250,60)[c]{V$_\pi$}
\BCirc(290,20){10}
\Text(290,20)[c]{$\Gamma_b$}
\end{picture}

\noindent Fig.~11. $\pi^0 \to\gamma\gamma$. There are also similar diagrams 
with $a \leftrightarrow b$.

\newpage

\begin{figure}
\center{\psfig{file=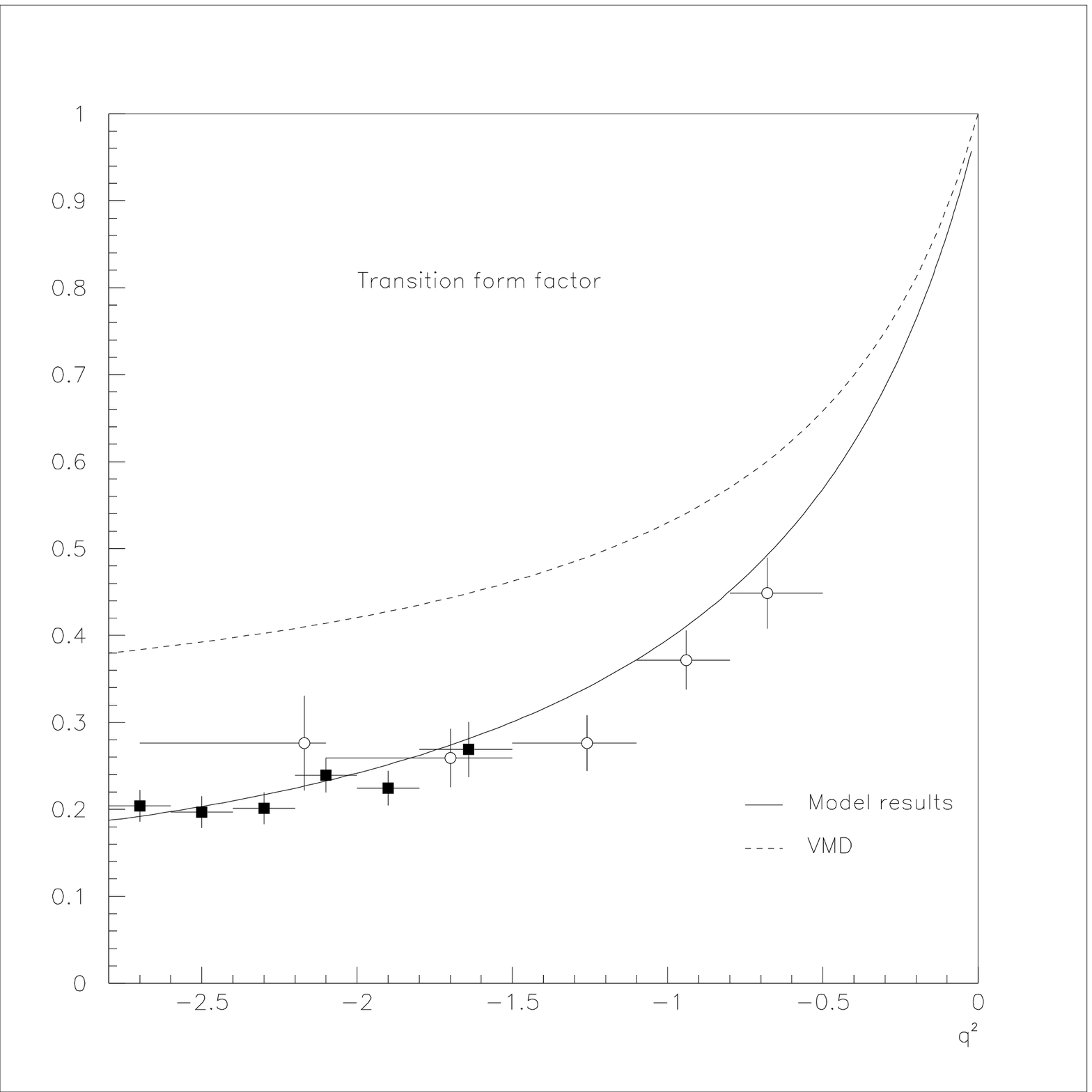,height=6in}}
\end{figure}

\noindent Fig.~12. The $\pi \gamma \gamma^\ast (q^2)$ transition form factor,
plotted against $q^2$ in GeV$^2$. The solid line is the model result; the
dashed line is the VMD approximation to it. The open data points are from 
Ref.\cite{CELLO}, the solid ones from Ref.~\cite{CLEO}.

\begin{figure}
\center{\psfig{file=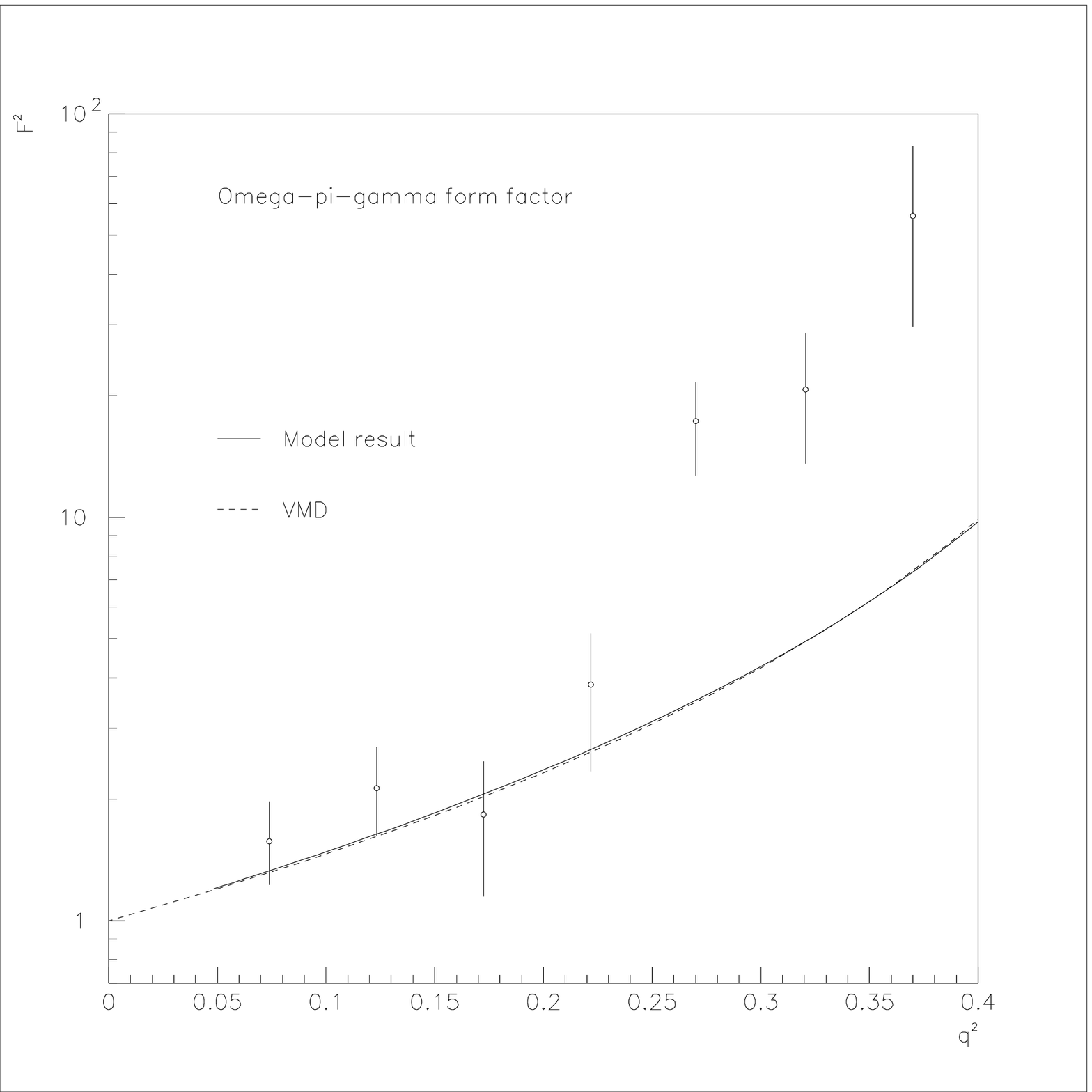,height=6in}}
\end{figure}

\noindent Fig.~13. The square of the $\omega \pi \gamma^\ast (q^2)$ form
factor, plotted on a logarithmic scale against $q^2$ in GeV$^2$. The solid
line is the model result; the dashed line is the VMD approximation to it.
The data points are from Ref.\cite{D+81}.

\end{document}